\documentclass[tightenlines,aps,twocolumn,
nofootinbib
]{revtex4}

\usepackage{psfig,float}
\usepackage{graphics}
\newcommand{\be}{\begin{equation}}
\newcommand{\ee}{\end{equation}}
\newcommand{\ba}{\begin{eqnarray}}
\newcommand{\ea}{\end{eqnarray}}

\def\ket#1{\vert #1 \rangle}
\def\bra#1{\langle #1 \vert}

\begin{document}
\begin{titlepage}

\begin{flushright}
\vbox{
\begin{tabular}{l}
UH-511-1041-03\\
 FTPI-MINN-03/36\\
 UMN-TH-2224/03\\
 hep-ph/0312226
\end{tabular}
}
\end{flushright}

\vspace{0.6cm}

\title{Hadronic light-by-light scattering contribution\\
to the muon anomalous magnetic moment revisited
}

\author{Kirill Melnikov \thanks{
e-mail:  kirill@phys.hawaii.edu}}
\affiliation{Department of Physics and Astronomy,\\ University of Hawaii,\\
Honolulu, HI, D96822}

\author{Arkady Vainshtein \thanks{
e-mail:  vainshte@umn.edu}}
\affiliation{William I. Fine Theoretical  Physics Institute, 
University of Minnesota,\\
116 Church St. SE, Minneapolis, MN 55455 \vspace{1.5cm} 
}

\begin{abstract}

\vspace{2mm}

We discuss hadronic light-by-light scattering contribution 
to the muon anomalous magnetic moment $a_\mu^{\rm lbl}$,
paying particular 
attention to the  consistent matching between the short- and 
the long-distance behavior of the light-by-light scattering amplitude. 
We argue that the short-distance 
QCD imposes strong constraints on this amplitude
overlooked in  previous analyses.
We find that accounting for these constraints 
leads to approximately $50\%$ increase in the central 
value of $a_\mu^{\rm lbl}$,
compared to existing estimates. 
The  hadronic light-by-light scattering contribution 
becomes 
$a_\mu^{\rm lbl} = 136(25) \times 10^{-11}$, 
thereby  shifting  the Standard Model prediction  
closer to the experimental value. 

\end{abstract}

\maketitle

\thispagestyle{empty}
\end{titlepage}

\section{Introduction}

Recent results \cite{g-2recent} 
from the experiment E821 at BNL
might indicate
a disagreement between the experimental value
of the muon  anomalous magnetic moment $a_\mu=(g_{\mu}-2)/2$
and the theoretical expectation based on   
the Standard Model (SM). 
Although  no definite conclusion is possible at the moment, the 
experimental value of $a_\mu$ 
is persistently higher than the SM prediction; the significance
of the deviation depends on subtle aspects of the low-energy 
hadronic physics.
The largest 
hadronic contribution to $a_{\mu}$ is due to 
vacuum polarization, see Fig.\,1a.
\begin{figure}[h]
\centerline{
\psfig{figure=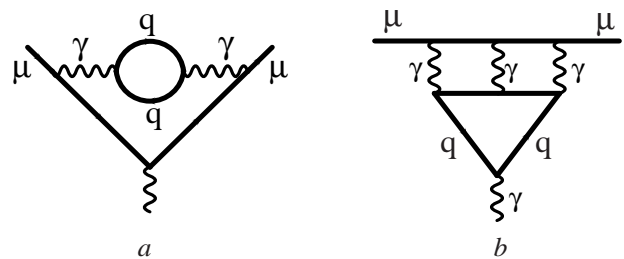,width=3.2in}
}
   \caption{Hadronic contributions represented by quark loops: 
($a$) vacuum polarization, ($b$) light-by-light scattering.}
\end{figure} 
It can be found by integrating  
the $\sigma(e^+e^- \to {\rm hadrons})$ 
annihilation cross-section with the weight function 
computed in perturbation theory.
Experimentally, the $e^+e^-$ annihilation cross section is obtained either
from direct  measurements at  low energies or, using the isospin 
symmetry,   from  hadronic decays of the $\tau$  meson.

The most recent study \cite{eidelman} 
 gives different results for 
the $e^{+}e^{-}$-- based and the $\tau$-- based analyses; the primary 
reason is  the  disagreement  between the $e^{+}e^{-}$ and the 
$\tau$ data in the energy range $0.85-1.0~{\rm GeV}$.  
It is this experimental issue
that currently limits precision in computing hadronic 
vacuum polarization contribution to $a_\mu$.
 
Another source of hadronic contributions to $a_{\mu}$ is  
the light-by-light scattering, induced by hadrons, see Fig.\,1b.
Compared to the  vacuum polarization, this contribution is 
significantly smaller; nevertheless, given experimental 
precision on $a_\mu$,  it is  quite  important.

The hadronic light-by-light scattering contribution 
cannot be related to experimental data; for this reason
the existing  estimates of this contribution 
are model dependent.   This feature leads to major problems 
in  estimating  both the central value and the theoretical uncertainty. 
Given the fact that at low energies the physics of light-by-light scattering 
is non-perturbative, it is na\"{\i}ve to expect the fully model-independent
solution. The satisfactory solution should involve a  mixture 
of both model-dependent and first-principles based considerations 
in such a way that the uncertainty caused by the model dependence 
can be  minimized and controlled.  

To quantify the quality of the low-energy hadronic model, we 
need a theoretical parameter. Since  the perturbation theory 
is not an option, we must look for the 
parameter other then the QCD coupling constant.
The two possibilities
are the smallness of the chiral symmetry breaking 
and the large number of colors $N_c$.  
The relevance of these parameters can be seen  from the 
parametrical expression for $a_{\mu}^{\rm lbl}$,
\begin{equation}
a_{\mu}^{\rm lbl}\sim 
\left ( \frac{\alpha}{\pi} \right )^{3} \left[ c_{1}\, \frac{m_{\mu}^{2}}{m_{\pi}^{2}} + c_{2} N_{c}\,\frac{m_{\mu}^{2}}{\Lambda_{\rm QCD}^{2}}\right],
\end{equation}
where it is assumed  that $m_{\pi}>m_{\mu}$. 
Only  the  power dependence on $m_{\pi}^{2}$ is shown; 
possible chiral logarithms  are included into the coefficients $c_{1,2}$. 
The first, chirally enhanced, term is due to the loops of charged pions 
in the light-by-light scattering, Fig.\,2a.
The second, $N_c$-enhanced, term is due to 
exchanges of neutral pion or heavier resonances, Fig.\,2b.\\[-0.8cm]
\begin{figure}[h]
\centerline{
\psfig{figure=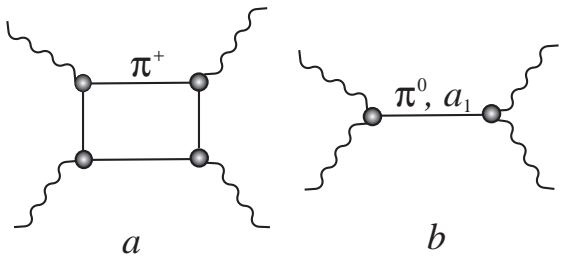,width=2.9in}}
   \caption{Hadronic contributions to the light-by-light scattering: 
($a$) charged pion loop, ($b$) exchange of neutral pion and other resonances.}
 \end{figure} 

At first sight, it seems natural to expect
the chiral  parameter $m_{\pi}^{2}/(4\pi f_{\pi})^{2}$ 
to be  a better expansion parameter for $a_\mu^{\rm lbl}$.
However,  a more careful analysis indicates that things can, and perhaps 
do, work differently. In particular,
in all hadronic models used to estimate $a_\mu^{\rm lbl}$,
the chirally enhanced two-pion contribution 
is always much smaller than the color enhanced contribution. We present 
the ``anatomy'' of the chirally enhanced ${\cal O}(N_c^0)$ contribution 
 in the last Section of this paper where we argue  that 
this smallness may not be accidental.

Moreover, a similar example  is provided by the 
hadronic vacuum polarization contribution to $a_\mu$. 
There, the chirally enhanced two-pion contribution gives 
approximately $3 \times 10^{-9}$ which should be 
compared with the $N_c$-enhanced contribution due to the $\rho$-meson 
that gives approximately $50 \times 10^{-9}$.  
Although we do not have a  clear understanding 
of why the chirally enhanced terms are subdominant to such an extent, 
the above arguments suggest that 
we should accept the dominance of the 
large-$N_c$ expansion over the chiral expansion 
as the working hypothesis.
The special feature  of the large-$N_c$ limit is that 
scattering amplitudes in any particular 
channel are given by  infinite sums of 
narrow resonances. This helps in constructing the model 
but is clearly insufficient; 
we need further constraints to select among  prospective models. 

Such  constraints come from the knowledge of   short-distance behavior
of  the light-by-light scattering amplitude, 
governed by QCD. The asymptotics of this amplitude
at large Euclidean photon momenta is derived 
from  the operator product expansion 
(OPE). 
The leading term in this OPE comes from the quark box diagram
enhanced  by large $N_c$. This shows a consistency of the OPE constraints 
with the  large-$N_{c}$ limit. 
Therefore, we  require  an  acceptable large-$N_c$ hadronic
model, extrapolated to large Euclidean photon momenta, to match 
the perturbative light-by-light scattering amplitude.
We find that  the {\it minimal} 
large-$N_c$ model which satisfies this criterion
includes exchanges of  the pseudoscalar 
$0^{-}$ mesons $\pi^{0}, \eta, \eta'$ and the  
 the pseudovector $1^{+}$ resonances $a_1, f_1, f_1^*$ .
It is important to emphasize at this point that 
the model with  a finite number of resonances  
is consistent with the short-distance constraints for $a_\mu^{\rm lbl}$;
it is known that this is  not always the case ( see \cite{BGLP} for a 
recent discussion).
 
The short-distance QCD constraints are most restrictive 
in the pseudoscalar isovector channel.
In a special kinematic limit, where invariant masses of two virtual photons
are much larger than the invariant mass of the third 
photon,   this channel is completely saturated  by 
the neutral pion. The saturation is complete in the sense that 
it works for arbitrary small invariant mass of the third virtual photon, 
in spite of the fact that, in general, the OPE applies only 
when that mass is much larger than $\Lambda_{\rm QCD}$.

This happens because in the  kinematic limit described above, 
the OPE relates hadronic light-by-light scattering 
diagram to the famous ``anomalous''  triangle diagram 
with one axial and two vector  currents.  
Because both perturbative and non-perturbative corrections 
to the anomalous triangle are absent in the limit of exact 
chiral symmetry, the pion pole contribution is unambiguous 
both at small $q \sim \Lambda_{\rm QCD}$ 
{\it and} at large $q \gg \Lambda_{\rm QCD}$ momenta.
This observation connects the two regions of momenta 
and provides an important constraint thereby. 

In terms of the diagram in  Fig.\,2a, the constraint amounts 
to the statement that the form factor
is present in the $\pi\gamma^{*}\gamma^{*}$
 vertex if both photons are virtual but 
it is {\em absent}  if that vertex contains the external magnetic field. 
Although the pseudoscalar channel has  been the subject of
many detailed studies in the past, this  constraint has been overlooked
and, as the result,  
the  $\pi^{0}$-pole contribution to $a_\mu$ was underestimated. 
This is the main source of corrections we 
find for the  pion pole contribution.

Moreover, additional  constraints on  subleading terms
 in  $\pi\gamma^{*}\gamma^{*}$  form factor, derived 
long ago in Ref.\cite{misuse}, were
not utilized previously. Accounting for these constraints,
also leads to the increase in the result.
As a consequence, the central value of the pion pole 
contribution to $a_\mu$ increases by approximately $20 \times 10^{-11}$.
Similar increases occur
for other pseudoscalar ($\eta,\eta'$) and 
pseudovector channels ($a_1,f_1,f_1^*$).

Unfortunately, the  constraints on all, but $\pi^{0}$,
exchanges  are not very restrictive; because of that
we cannot claim  significant reduction in the theoretical uncertainty 
of hadronic light-by-light scattering 
contribution to $a_\mu$.  Nevertheless, imposing 
all the constraints  from the short-distance QCD, we arrive 
at $a_\mu^{\rm lbl} = 136(25) \times ^{-11}$ which is approximately
$50$ per cent larger than  the existing estimates 
\cite{bijnes,klbl,klbl2,knecht}.

The rest of the paper is organized as follows. In the next Section 
we discuss the constraints coming from the short-distance
QCD and  a minimal model for 
hadronic contributions to $a_\mu^{\rm lbl}$.
We  consider the pseudoscalar 
and the pseudovector exchanges in Sections III and IV, respectively. 
In Section V we  briefly 
discuss the ${\cal O}(N_c^0)$ pion box contribution 
to $a_\mu^{\rm lbl}$. 
We present our conclusions in Section VI. Additional 
formulas are given in Appendices.

\section{Short-distance QCD constraints and hadronic model}

In this Section we describe  the constraints coming 
from the short-distance 
QCD and formulate the hadronic model that satisfies  these constraints.
\subsection{Kinematics}
 We begin  with the kinematics. 
The light-by-light scattering amplitude 
involves four photons with momenta $q_i$ 
and the polarization vectors $\epsilon_{i}$. 
We take the photon momenta to be incoming, $\sum q_{i}=0$.
The first three photons 
are virtual, while the fourth one  represents the external magnetic field 
and can be regarded as a real photon with 
the vanishingly small momentum $q_{4}$. The amplitude ${\cal M}$ is 
defined as
\begin{widetext}
\ba
{\cal M}&=&\alpha^{2}N_{c}\,{\rm Tr} \,[{\hat Q}^{4}] \,{\cal A}=
\alpha^{2}N_{c}\,{\rm Tr} \,[{\hat Q}^{4}] \,{\cal A}_{\mu_{1}\mu_{2}\mu_{3}\gamma\delta}\epsilon_{1}^{\mu_{1}} \epsilon_2^{\mu_2}\epsilon_3^{\mu_3}f^{\gamma\delta}
\nonumber \\[1mm]
&=&
- e^{3}\! \int\!  {\rm d}^{4}x\, {\rm d}^{4}y \,{\rm e}^{-iq_{1}x-iq_{2}y}\,
\epsilon_{1}^{\mu_{1}} \epsilon_2^{\mu_2}\epsilon_3^{\mu_3}
 \bra{0}
T\left\{j_{\mu_{1}}(x)\, j_{\mu_{2}}(y)\,j_{\mu_{3}}(0)\right\}\ket{\gamma},
\label{eqcala}
\ea
\end{widetext}
where $j_{\mu}$ is the hadronic electromagnetic current, 
$j_{\mu}\!=\!\bar q \,\hat Q \gamma_{\mu} q$, written
in terms of  the three quark 
flavors $q\!=\!\{u,d,s\}$ with $\hat Q$ being  the $3\times 3$ diagonal 
matrix of quark electric charges. In addition, 
$f^{\gamma\delta}\!=\!q_{4}^{\gamma}\epsilon_4^{\delta}-q_{4}^{\delta}\epsilon_4^{\gamma}$ denotes the field strength tensor of the soft photon; 
the light-by-light scattering amplitude  is proportional to this 
tensor due to gauge invariance. Since  ${\cal M}$ is linear in 
the small momentum $q_{4}$, for the purpose of computing 
the light-by-light scattering contribution to $a_\mu$,
we can set $q_4=0$ in  the tensor amplitude 
${\cal A}_{\mu_{1}\mu_{2}\mu_{3}\gamma\delta}$ 
and calculate it assuming that  $q_1 + q_2 + q_3 = 0$ 
for the virtual photons. Because the momenta $q_{1}, q_{2}, q_{3}$ form a 
triangle, there are just three independent 
Lorentz invariant variables;  we choose them to be
the virtualities of the photons 
$q_{1-3}^{2}$.

In general, the light-by-light scattering amplitude 
is  a complicated function 
of photon's virtualities. However, there are only two distinct kinematic  regimes in the 
light-by-light scattering amplitudes: the Euclidean momenta of the 
three photons are comparable in magnitude
$q_{1}^{2} \sim q_{2}^{2}\sim q_{3}^{2}$, or 
one of the momenta is much smaller than the other two.
The second limit can be analyzed in a very simple fashion 
using the OPE of the light-by-light scattering. 
Also, this limit is of importance because 
it helps us to identify 
the pole-like structures in the OPE amplitudes  
and in this way connect the OPE to phenomenological models.

\subsection{OPE and triangle amplitude}
Since the light-by-light  scattering  amplitude is symmetric
with respect to photon permutations, we can study the 
second limit assuming that 
$q_1^{2} \approx q_2^{2}\gg q_3^{2}\,$. 
In this kinematic regime, we begin with 
the well-known OPE  (see e.g.  \cite{Bj}) for
the product of two electromagnetic currents 
that carry  the largest momenta $q_{1}, q_{2}$, 
\begin{eqnarray}
\label{jj}
&&i\!  \int\!  {\rm d}^{4}x\, {\rm d}^{4}y\,{\rm e}^{-iq_{1}x-iq_{2}y}\, 
T\left\{j_{\mu_{1}}(x),j_{\mu_{2}}(y)\right\}=
\nonumber \\[1mm]
&&\int\!  {\rm d}^{4}z \,
{\rm e}^{-i(q_{1}+q_{2})z} \, \frac{2i}{{\hat q}^{2}}\,
\epsilon_{\mu_{1}\mu_{2}\delta\rho}\,{\hat  q}^{\delta}j_{5}^{\rho}(z)
+\cdots\,. 
\end{eqnarray}
Here, 
$j^{\rho}_{5}=\bar q\, \hat Q^{2} \gamma^{\rho} \gamma_{5}\,q$  is 
the axial current, where different flavors enter
 with weights proportional to squares of their 
electric charges and 
$\hat q=(q_{1}-q_{2})/2\approx q_{1}\approx -q_{2}\,$.  We retain 
only the leading (in the limit of large Euclidean $\hat q$) term in 
the OPE associated with the axial current $j_{5}^{\rho}$; 
the ellipsis in Eq.(\ref{jj}) stands 
for subleading terms suppressed by powers of $\Lambda_{\rm QCD}/\hat q$. 
The momentum $q_{1}+q_{2}=-q_{3}$ flowing 
through $j^{5}_{\rho}$ is assumed to be much smaller than $\hat q$.
We note in passing that Eq.(\ref{jj}) has been applied earlier 
in various situations; for example, 
the matrix element of Eq.(\ref{jj}) between the pion and the  
vacuum states gives the asymptotic behavior of  
the $\pi^{0}\gamma^{*}\gamma^{*}$ amplitude at large photon 
virtualities \cite{misuse}.

For the purpose of further discussion it is convenient to present the current 
$j_{5\rho}$ as a linear combination of the isovector, 
$j_{5\rho}^{(3)}=\bar q \lambda_{3}\gamma_{\rho}\gamma_{5}q$,
hypercharge, $j_{5\rho}^{(8)}=\bar q \lambda_{8}\gamma_{\rho}\gamma_{5}q$, 
and the SU(3) singlet,
$j_{5\rho}^{(0)}=\bar q \gamma_{\rho}\gamma_{5}q$, currents,
\be
\label{expj}
j_{5\rho}=\sum_{a=3,8,0}\frac{{\rm Tr}\,[\lambda_{a} \hat Q^{2}]}{{\rm Tr}\,[\lambda_{a}^{2}]}\,j_{5\rho}^{(a)}\,,
\ee
where $\lambda_{0}$ is the unity matrix.  

Once the dependence on the largest momenta $q_{1,2}$ is 
factored out,  the next step is 
to find the dependence of the light-by-light scattering 
amplitude  on the momentum $q_{3}$.  
This dependence is given by the amplitudes $T_{\gamma\rho}^{(a)}$ that
involve  axial currents $j_{5\rho}^{(a)}$ 
and two electromagnetic currents, one with momentum $q_3$ and 
the other one (the external magnetic field) with the  vanishing momentum
\ba
\label{trian}
T_{\mu_{3}\rho}^{(a)}
\!=i\,\langle 0| \!\int\! {\rm d}^4 z\, {\rm e}^{iq_{3}z}T\{ j^{(a)}_{5\rho} (z)\,j_{\mu_{3}} (0)\}|\gamma \rangle.
\ea

The triangle amplitudes for such kinematics
were considered  recently in  \cite{CMV}.
It is shown in that reference that 
$T_{\gamma\rho}^{(a)}$ can be written 
through two independent amplitudes,
\ba
&& T_{\mu_{3}\rho}^{(a)}=-\frac{ie\,N_{c}{\rm Tr}\,[\lambda_{a} \hat Q^{2}]}{4\pi^{2}}
\left\{w^{(a)}_L\!(q_{3}^2)\, 
q_{3\rho} q_{3}^\sigma  \tilde f_{\sigma  \mu_{3}}+
\right. \nonumber \\ 
&& \left. + w^{(a)}_T\!(q_{3}^2) \!\left(-q_{3}^2 \tilde f_{\mu_{3}\rho}\!+\!q_{3\mu_{3}} q_{3}^\sigma
\tilde f_{\sigma  \rho}\! -\! q_{3\rho} q_{3}^\sigma \tilde f_{\sigma  \mu_{3}
}\right)\right\}.
\label{eqw}
\ea
The first (second) amplitude is related to the longitudinal (transversal) 
part of the axial current, respectively. 
In terms of hadrons, the  invariant function $w_{L(T)}$ 
describes the  exchanges of the pseudoscalar (pseudovector) mesons.

In perturbation theory $w_{L,T}$ are defined by the famous triangle diagram. 
For massless quarks, we obtain:
\be
w^{(a)}_{L}(q^{2})=2w^{(a)}_{T}(q^{2})=-
\frac{2}{q^2}\,.
\label{eqLT}
\ee
An appearance of the longitudinal part is the  signature  
of the axial Adler-Bell-Jackiw (ABJ) anomaly \cite{ABJ}.
Although the perturbation theory is only reliable for 
$q^{2}\gg \Lambda_{\rm QCD}^{2}$, where it coincides with the 
leading term of the OPE for the time-ordered product of the axial and 
electromagnetic currents, the expressions for longitudinal 
functions $w^{(3,8)}_{L}$ given in Eq.(\ref{eqLT}) are 
{\em exact} QCD results in the chiral limit $m_{q}=0$ 
for nonsinglet axial currents. 
The fact that there are no
perturbative \cite{adler} 
and nonperturbative \cite{thooft} corrections to the axial anomaly 
implies that the pole behavior of  $w^{(3,8)}_{L}$ in Eq.(\ref{eqLT})
is correct all the way down to small $q^{2}$, 
where the poles are associated  with  Goldstone 
pseudoscalar mesons,  $\pi^{0}$ 
in $w^{(3)}_{L}$ and  $\eta$   in $w^{(8)}_{L}$.

Equations (\ref{eqw}) and (\ref{eqLT}) allow us to derive the coupling 
of the $\pi^{0}$ meson to photons. To this end,
consider the isovector part of the triangle 
amplitude $T_{\gamma\rho}^{(3)}$.   The residue at $q^2=0$, 
corresponding to the $\pi^{0}$ pole, 
is the product of two matrix elements,
\ba
\langle 0| j^{(3)}_{5\rho}|\pi^{0}\rangle  =2iF_{\pi}\,q_{\rho}  \,,
~~~\langle \pi^{0}| j_{\mu_{3}}|
\gamma\rangle =-4e g_{\pi 
\gamma\gamma}q^{\sigma}\tilde f_{\sigma\mu_{3}}\,.
\ea
Comparing with Eqs.(\ref{eqw}), (\ref{eqLT}) we derive  
the well-known result \cite{ABJ} for $\pi\gamma\gamma$ coupling: 
\be
g_{\pi \gamma\gamma}=\frac{N_{c}{\rm Tr}\,[\lambda_{3}{\hat Q}^{2}]}
{16\pi^{2} \, F_{\pi}}\,.
\label{gpgg}
\ee
In a similar way, the $g_{\eta \gamma\gamma}$ 
coupling in the chiral limit can be derived, if needed.

The absence of perturbative and non-perturbative 
corrections and therefore the possibility to use the OPE 
expressions for vanishing values of $q^2$ 
is unique for the longitudinal part 
of  nonsinglet axial currents.\footnote{More precisely, 
perturbative corrections to $w_{T}^{(3,8)}$ 
are also absent as  shown in Ref.\cite{nonren}.}
For the transversal functions $w_{T}$ as well as 
for the singlet longitudinal function  $w_{L}^{(0)}$, 
there are higher order  terms in the OPE that, 
upon summation, generate  mass terms that 
shift the pole position $1/q^2 \to 1/(q^2-m^2)$.
We  use this modification of the pole-like terms 
for each channel in what follows.
The lightest pseudovector mesons  are the $a_1(1260)$,
$f_1(1290)$  and $f_1^{*}(1420)$ mesons.  
For the singlet axial current the pole in $w_{L}^{(0)}$ 
is shifted to $m_{\eta'}^{2}\,$. 

Consider a triangle amplitude for any  isospin channel 
in the limit of large $q^{2}$, where the OPE and the perturbation theory 
are applicable and  Eq.(\ref{eqLT}) is valid. 
An important  consequence of this equation is that triangle 
amplitudes are {\em not} suppressed for  such values of 
$q^{2}$. 
In terms of hadrons it means that no 
form factor is present  in 
the $h \gamma^* \gamma$ interaction  vertex where 
the real photon   is soft (external magnetic field). 
This is in  clear contradiction with the common 
practice \cite{klbl,bijnes,knecht} when,  
for $\pi^{0}$ exchange,  the form factor 
$F_{\pi \gamma^* \gamma}(q^2,0)$ 
is introduced. 
Such transition form factor has to be 
present when one of the photons  is virtual, the other  photon is 
on the mass shell and  the pion is on the mass shell as well. 
However, this is not the  kinematics that corresponds to the triangle and 
the light-by-light scattering  amplitudes, 
relevant for $a_\mu$ computation, where 
the pion virtuality is {\em the same} as the virtuality  of one of 
the photons.
The  absence of the suppression is also consistent with the 
dispersion representation of the amplitude since  
the imaginary part is nonvanishing only at $q^{2}=0$ 
in the chiral limit.

\newpage

\begin{widetext}
Combining Eqs.(\ref{jj}-\ref{eqw}), we write the 
the  light-by-light amplitude ${\cal A}_{\mu_{1}\mu_{2}\mu_{3}\gamma\delta}$
for  $q_{1}^{2}\approx q_{2}^{2}\gg q_{3}^{2}$ in the following form:
\ba
{\cal A}_{\mu_{1}\mu_{2}\mu_{3}\gamma\delta} f^{\gamma\delta}=
\frac{8}{{\hat q}^{2}}\epsilon_{\mu_{1}\mu_{2}\delta\rho}{\hat  q}^{\delta}
\!\!\sum_{a=3,8,0}\!\! W^{(a)}
\left\{w^{(a)}_L\!(q_{3}^2)\, q_{3}^{\rho} q_{3}^\sigma  \tilde f_{\sigma  \mu_{3}}+
w^{(a)}_T\!(q_{3}^2) \!\left(-q_{3}^2 \tilde f_{\mu_{3}}^{\rho}\!+\!q_{3\mu_{3}} q_{3}^\sigma
\tilde f_{\sigma}^{\rho}\! -\! q_{3}^{\rho} q_{3}^\sigma \tilde f_{\sigma  \mu_{3}
}\right)\right\}
+ \cdots,
\label{eq5}
\ea
\end{widetext}
where no hierarchy between $q_3^2$ and $\Lambda_{\rm QCD}^2$  
is assumed. The weights $W^{(a)}$ are defined as 
\ba
\label{weights}
&& W^{(a)}=  \frac{\left ( {\rm Tr}\,[\lambda_{a} \hat Q^{2}] \right )^2}{{\rm Tr}\,[\lambda_{a}^{2}]{\rm Tr}\,[\hat Q^{4}]}\,;
\\[1mm]
&& W^{(3)}=\frac{1}{4}\,,\quad W^{(8)}=\frac{1}{12}\,,
\quad  W^{(0)}=\frac{2}{3}\,.
\nonumber 
\ea

In the limit  $q_{3}^2\gg\Lambda_{\rm QCD}^{2}$, Eq.(\ref{eq5})
can be simplified using 
 the asymptotic expressions Eq.(\ref{eqLT}) for 
the invariant functions $w_{L,T}^{(a)}$. 
Convoluting the tensor amplitude 
with the photon polarization vectors 
and analytically continuing to Euclidean space, we arrive at:
\ba
&& {\cal A}\!=\!\frac{4}{q_3^2 \,\hat q^2} \{f_2 \tilde f_1 \} \{\tilde f f_3\}
\!-\!\frac{4}{q_3^2 \,\hat q^4}\!\left(\!\!\{q_{2}f_{2}\tilde f_{1} \tilde f f_{3}q_{3}\}\! \right.
\nonumber \\
&& \left. +\!\{q_{1}f_{1}\tilde f_{2} \tilde f f_{3}q_{3}\}
\!+\! \frac{q_{1}^{2}+q_{2}^{2}}{4}\{f_{2}\tilde f_{1}\}\{\tilde f f_{3}\}\!\!\right)\!+\! \cdots\!. 
\label{amplpole}
\ea
Here,  $f_{i}^{\mu\nu}=q_{i}^{\mu}\epsilon_{i}^{\nu}
- q_{i}^{\nu}\epsilon_{i}^{\mu}$ are the field strength tensors,  
the  braces denote  either traces of products of the 
matrices $f_{i}^{\mu\nu}$
or their convolutions with vectors $q_{i}$.

In Eq.(\ref{amplpole}) and in the remainder of the 
paper, we use Euclidean notations 
instead of Minkowski ones used before. 
The continuation to Euclidean space mostly concerns the change in sign 
for all $q_i^2$ and the overall 
change in sign for the amplitude ${\cal A}$, since it involves the 
product of two Levi-Cevita tensors.
The result can be verified by comparison with the direct
computation of the quark box diagram, for arbitrary $q_{1-3}^{2}$,
presented in Appendix I.
There we show that the amplitude can be described in terms of nineteen 
independent tensor structures and five independent form-factors.
In what  follows, we  mostly deal with the approximate 
form of the amplitude  Eq.(\ref{amplpole}), but we  make 
occasional references to general expression in Appendix I.

\subsection{The model}

Two different terms in 
Eq.(\ref{eq5}) can be identified with  exchanges 
of the pseudoscalar (pseudovector) mesons 
for the  functions $w^{(a)}_{L,T}(q^{2}_{3})$. 
Extrapolating Eq.(\ref{amplpole}) from 
$q^{2}_{1,2} \gg \Lambda_{\rm QCD}^2$ to arbitrary $q^{2}_{1,2}$, 
we arrive at the following model:
\be
{\cal A}\!=\!{\cal A}_{\rm PS}+{\cal A}_{\rm PV}\, 
+{\rm permutations},\\[2mm]
\ee
where
\begin{widetext}
\ba
{\cal A}_{\rm PS}&\!=&\!\!\!\sum_{a=3,8,0}\!\! W^{(a)} \phi^{(a)}_{L}(q_1^{2},q_2^{2})
w_L^{(a)}(q_3^{2})
\{f_{2}\tilde f_{1}\}\{\tilde f f_{3}\}
,\label{PS}\\[1mm]
{\cal A}_{\rm PV}&\!=&\!\!\!\sum_{a=3,8,0}\!\! W^{(a)} \phi^{(a)}_{T}(q_1^{2},q_2^{2})
w_T^{(a)}(q_3^{2})
\left(\!\{q_{2}f_{2}\tilde f_{1} \tilde f f_{3}q_{3}\}
\!+\!\{q_{1}f_{1}\tilde f_{2} \tilde f f_{3}q_{3}\}
\!+\! \frac{q_{1}^{2}+q_{2}^{2}}{4}\{f_{2}\tilde f_{1}\}\{\tilde f f_{3}\}\!\right). 
\label{PV}
\ea
\end{widetext}
The form factors $\phi^{(a)}_{L,T}(q_1^{2},q_2^{2})$ 
account for the dependence of the amplitude on $q_{1,2}^{2}$. 
Pictorially (see Fig.2b), these form factors can be associated with
the interaction vertex for the two virtual photons 
on the left hand side, whereas 
the meson propagator and  the interaction vertex on the right hand side
form the triangle amplitude described 
by the functions $w_{L,T}^{(a)}(q_{3}^{2})$. 
In the next Sections we introduce models for these functions  consistent 
with the short distance behavior of the light-by-light scattering 
amplitude. 

Note that our model  does not 
include explicit exchanges of vector or scalar mesons.  
This is a  consequence of the fact that, to leading order,  the 
OPE of the two vector currents 
produces the axial vector current only. 
However, the vector mesons are 
present in our model implicitly,  through the momentum 
dependence of  the form factors $\phi^{(a)}_{L,T}$ 
as well as the transversal  functions $w^{(a)}_{T}$.

\section{Constraints on the pseudoscalar exchange}

The $\pi^{0}$ exchange 
provides the largest fraction of the  hadronic light-by-light 
scattering contribution to $a_\mu$. It is therefore 
appropriate to scrutinize this contribution as much as 
possible and ensure that it satisfies 
all the possible constraints that follow from first principles.

As we discussed earlier, 
the longitudinal part of the triangle amplitude is fixed by the ABJ anomaly.
Accounting for explicit violation of the chiral symmetry given by 
the small mass of the pion, we derive
\be
w_{L}^{(3)}(q^{2})=\frac{2}{q^{2}+m_{\pi}^{2}}\,.
\ee
The  ABJ anomaly also fixes  $\phi_{L}^{(3)}(0,0)$,
\be
\phi_{L}^{(3)}(0,0)=\frac{N_{c}}{4\pi^{2}F_{\pi}^{2}}\,,
\ee
so that the model for the pion exchange in 
the light-by-light scattering amplitude takes the form,
\ba
{\cal A}_{\pi^{0}}&\!=&\!-\frac{N_{c}W^{(3)}}{2\pi^{2}F_{\pi}^{2}}\, \frac{F_{\pi \gamma^* \gamma^*}(q_1^2,q_2^2) }{q_{3}^{2}+m_{\pi}^{2}}\,\{f_{2}\tilde f_{1}\}\{\tilde f f_{3}\}
\nonumber\\[1mm]
&&+ \,\mbox{permutations}\,.
\label{pionmod}
\ea
The $\pi \gamma^* \gamma^*$  form factor $F_{\pi \gamma^* \gamma^*}(q_1^2,q_2^2)$ is defined 
as 
\be 
F_{\pi \gamma^* \gamma^*}(q_1^2,q_2^2)=\frac{\phi_{L}^{(3)}(q_{1}^{2},q_{2}^{2})}{\phi_{L}^{(3)}(0,0)}\,.
\ee
The comparison with the OPE constraint given by the relevant term in 
Eq.(\ref{eq5}) 
leads to 
\be
\lim_{q^{2} \gg \Lambda_{\rm QCD}^{2}} 
F_{\pi \gamma^* \gamma^* }(q^2,q^2) = \frac{8 \pi^2 F_\pi^2}{N_c\, q^2}\,,
\label{1q2}
\ee
which is the correct asymptotics indeed \cite{misuse}. 
This means that the neutral pion  exchange in   
Eq.(\ref{PS}) saturates the 
corresponding short-distance QCD constraint.

This comparison also proves  our previous claim that the  
 form factor  $F_{\pi \gamma^* \gamma }(q_{3}^2,0)$ 
cannot be present in the amplitude Eq.(\ref{pionmod});
if that form factor is introduced,  the asymptotics 
of the light-by-light scattering amplitude becomes $1/q_3^4$, 
as opposed to $1/q_3^2$ behavior that follows from perturbative 
QCD.  This proof is, of course, equivalent 
to our discussion of the triangle amplitude in Section II.

The absence of the second form factor in the amplitude 
Eq.(\ref{pionmod})
distinguishes our approach from 
all other calculations of the pion pole 
contribution to  $a_\mu$ that exist in the literature.
As we show below, it has a non-negligible impact on the final numerical 
result for the pseudoscalar contribution to $a_\mu$. 
Here we note, that  the result for the pion pole contribution 
is expected  to
{\it increase}, because the absence of 
 the second form factor leads to  slower convergence 
of the integrals over loop momenta, making the result  larger.
As we show below, this is indeed what happens.

Further constraints on the model follow from   
restrictions on the pion transition form factor $F_{\pi\gamma^{*}\gamma^{*}}$
that were recently reviewed in \cite{knecht}. 
For numerical estimates we use their LMD+V form factor
\begin{widetext}
\be
F_{\pi \gamma^* \gamma^*}(q_1^2,q_2^2)
= \frac{4\pi^2 F_\pi^2}{N_c}~ 
\frac{q_1^2q_2^2(q_1^2+q_2^2)
-h_2 q_1^2 q_2^2 + h_5 (q_1^2+q_2^2)+(N_cM_1^4M_2^4/4\pi^2 F_\pi^2)
}{(q_1^2 +M_1^2)(q_1^2+M_2^2)(q_2^2+M_1^2)(q_2^2+M_2^2)},
\label{lmdv}
\ee
\end{widetext}
where $M_1 =769~{\rm MeV}$, $M_2 = 1465~{\rm MeV}$,  $h_5 =6.93~{\rm GeV}^4$.

The parameter $h_2$ was not determined in Ref.\cite{knecht} and 
we can fix it if we notice that
it contributes to the $1/q^4$ correction to the leading asymptotics
of the pion form factor, 
Eq.(\ref{1q2}). 
Such correction comes from the twist 4 operators in
 the OPE expansion of the two electromagnetic currents 
Eq.(\ref{jj}). It was analyzed 
long ago in Ref.\cite{misuse} using  the OPE and the QCD sum rules 
approaches.
The  result of such an analysis implies that 
the coefficient of the ${\cal O}(q^{-4})$ term in the asymptotics of the pion 
form factor is numerically small; 
in terms of the parametrization Eq.(\ref{lmdv}), this 
means that  $h_2 \approx -10~{\rm GeV}^2$ has to be chosen.  
We  use this value for  numerical estimates in what follows. 

Equations (\ref{pionmod}) and (\ref{lmdv}) completely specify the model
for the pion  pole contribution that we use for numerical 
calculations below. Before going into  that, we discuss the sensitivity 
of the final result to possible modifications of the model.

We denote  the structure that multiplies 
$\{f_{2}\tilde f_{1}\}\{\tilde f f_{3}\}$ in Eq.(\ref{pionmod})
as $W^{(3)} G_2^{\rm mod}(q_3^2,q_2^2,q_1^2)$. Comparing the 
$\pi^0$-pole exchange amplitude,  Eq.(\ref{pionmod}), to 
the full light-by-light scattering amplitude (see Appendix I), 
we find that, for asymptotically large virtualities of the photons,
the matching requires
\be
 G_2^{\rm mod}(q_3^2,q_2^2,q_1^2) = 
 G_2(q_3^2,q_2^2,q_1^2).
\label{gcomp}
\ee

Consider Eq.(\ref{gcomp}) in the limit
$\Lambda_{\rm QCD}^{2} \ll q_1^{2} \ll q_2^{2} \sim q_3^{2}$. 
It is easy to see that the left hand side
in Eq.(\ref{gcomp}) develops the $1/q_1^2$ behavior; 
from expression for $G_2$ in Appendix I it follows
that $G_2(q_2^2,q_2^2,q_1^2) \approx 1/q_2^2$ in such 
kinematic regime. Hence, there is a mismatch between our model 
and the OPE prediction.
 
The second option  is to consider Eq.(\ref{gcomp})
in the situation when all the momenta 
are asymptotically large and equal in magnitude
$q_1^{2}=q_2^{2}=q_3^{2}=q^{2}$. In this regime, 
\be
 G_2(q^2,q^2,q^2) \approx \frac{8}{3q^4},
\label{eqas}
\ee
whereas 
\be
G_2^{\rm mod} (q^2,q^2,q^2) \approx \frac{4}{q^4}.
\ee
Again, the model fails to describe the OPE constraint perfectly.

Of course, the above failures do not necessary invalidate the 
model; after all 
we are interested in the light-by-light scattering contribution 
to $a_\mu$ and various regions of loop momenta 
contribute differently to the integral.  
For this reason, we have to investigate if 
the above mismatches influence the 
numerical estimate for the pion pole contribution to 
$a_\mu^{\rm lbl}$. To this end, we notice that
$G_2^{\rm mod} (q_3^2,q_2^2,q_1^2)$ can be modified 
by adding to it 
\ba
&&\hspace{-10mm}\delta G_2^{\rm mod} (q_3^2,q_2^2,q_1^2) 
=
\nonumber\\
&&\frac{ \xi  q_1^2 q_2^2}{
(q_1^2 +M_1^2)(q_1^2+M_2^2)(q_2^2+M_1^2)(q_2^2+M_2^2)},
\ea
without running into a contradiction with 
the required pole behavior with respect to  $q_3^2$. 
After adding $\delta G_2^{\rm mod}$, it is easy to 
see that, by tuning $\xi$, one can either ensure 
that the pole in $q_1^2$ is absent or that 
the asymptotic behavior of $G_2^{\rm mod}+\delta G_2^{\rm mod}$
becomes consistent with Eq.(\ref{eqas}).
The two constraints are satisfied
for $\xi = -2$ and $\xi = -4/3$, respectively.

We can investigate the importance of these constraints 
by computing the contribution of $\delta G_2^{\rm mod}$
to $a_\mu^{\rm lbl}$ for $\xi=1$. Upon doing so, we find that
it changes $a_\mu^{\pi^{0}}$
by $\approx 0.5 \times 10^{-11}$. Hence, regardless 
of the value of $\xi$,   $\delta G_2^{\rm mod}$
can be neglected at the current level of precision. 
We therefore use Eqs.(\ref{pionmod},\ref{lmdv}) as our model 
for the pion form factor in what follows.

The result for $a_\mu^{\pi^0}$ with the 
LMD+V form factor for $h_2 = -10~{\rm GeV}^2$, 
quoted in \cite{knecht} is $a_\mu^{\rm \pi^0} = 63 \times 10^{-11}$.  
Using the formulas in \cite{knecht} it is easy to repeat 
their calculation removing the pion transition 
form factor that involves the soft photon.
In that case, the result becomes $ 76.5~ \times 10^{-11}$,
a shift in the positive direction. In addition, as we mentioned 
earlier, we consider the value $h_2 = -10~{\rm GeV}^2$ 
to be preferable because of the OPE constraints 
on   the pion transition form factor. 
Note, however, that $h_2 = 0$ 
was used in \cite{knecht} to derive the central value 
$58 \times 10^{-11}$; compared to that number, 
our central value  is larger by approximately $20 \times 10^{-11}$.

A similar analysis for the isosinglet channels leads to the 
 conclusion that these channels are saturated by $\eta$ and 
$\eta'$ mesons;  matching to pQCD result suggests that 
no transition form factor is present for the soft photon 
interaction vertex in those cases as well. 
Since these contributions are smaller 
than that of $\pi_0$, we do not use  sophisticated models for 
$\eta$ and $\eta'$ transition form factors and 
estimate them using the 
simplest possible VMD form factor.\footnote{The VMD form factor 
obviously violates the scaling of the form factor when both 
photon virtualities become large. We have checked that using 
the form factor consistent with the asymptotic scaling $1/q^2$
at large  values of $q_{1,2}$ has no bearing on the final result for 
the $\eta$ and $\eta'$ contributions.}
The $\eta(\eta') \gamma^* \gamma^*$ interaction vertex is normalized
in such a way that  the decay widths of these mesons into two photons
is correctly reproduced; this allows to account for the 
$\eta-\eta'$ mixing in a simple way. 

How good these ``experimental'' couplings compare to the theoretical 
expectations based on our model? Because of the 
$\eta-\eta'$ mixing, we expect that the sum of $\eta(\eta')\gamma \gamma$ 
couplings squared is predicted by the model more accurately
than each of the couplings separately.
We find 
\be
r = \frac{ g_{\eta \gamma \gamma} ^2 
+g_{\eta' \gamma \gamma} ^2  }{g_{\pi \gamma \gamma} ^2} = 3,
\ee
whereas using experimental values for the $\eta(\eta')\gamma \gamma$ 
couplings we arrive at $r = 2.5(1)$. 
Although we  use this $20\%$ discrepancy 
as an error estimate on the $\eta + \eta'$ 
contribution, we note that it  rather implies  {\it an increase} 
in the result since the agreement between ``experimental'' and
theoretical asymptotics can be improved by 
adding more  pseudoscalar mesons to the model.

Compared to the results quoted in \cite{knecht}, 
removal of the second form factor increases the  $\eta$ 
and $\eta'$ contributions from approximately 
$a_{\mu}^{\eta} = a_\mu^{\eta'} = 13 \times 10^{-11}$ 
to $ a_{\mu}^{\eta} = a_\mu^{\eta'} = 
18 \times 10^{-11}$.  
The sum of the contributions from all  
pseudoscalar mesons ($\pi^{0},\eta,\eta'$) 
leads to the  estimate:
\be
a_{\mu}^{\rm PS} = 114(10) \times 10^{-11}.
\label{eq13f}
\ee

The central value in Eq.(\ref{eq13f})
is almost 40  percent 
larger than most of the existing 
results for $a_\mu^{\rm lbl}$
\cite{klbl,bijnes,knecht}. 
The major effect comes  from removing the form factor 
for the interaction of the {\it soft} photon
(magnetic field)  with the pseudoscalar meson; the necessity 
to do that {\em unambiguously} follows from matching the 
pseudoscalar pole amplitude to the pQCD expression for the
light-by-light scattering.

On the contrary, the error estimate in Eq.(\ref{eq13f}) is  
subjective; it is based on the variation of the result when input 
parameters of the model are varied. It is impossible to 
defend the {\it exact} number for the error estimate in Eq.(\ref{eq13f});
however, we believe that it adequately describes our current knowledge 
of the pseudoscalar contribution.

\section{Pseudovector exchange}
In this Section we discuss the pseudovector exchange amplitude 
${\cal A}_{\rm PV}$,  Eq.(\ref{PV}). From 
Eqs. (\ref{eqLT}), (\ref{eq5}), (\ref{amplpole}), we  find
the  asymptotics of $\phi^{(a)}_{T}$ and $w^{(a)}_{T}$,
\be
\hspace*{-0.5cm}
\lim_{\quad q^{2}\gg \Lambda_{\rm QCD}^{2}}\hspace{-3mm}\phi^{(a)}_{T}(q^{2}, q^{2})= \frac{-4}{q^{4}},
\lim_{\quad q^{2}\gg \Lambda_{\rm QCD}^{2}}\hspace{-3mm} w^{(a)}_{T}(q^{2})= \frac{1}{q^{2}}.
\label{Tcon}
\ee
As we mentioned earlier,
the lightest pseudovector resonances 
are  the $a_1$ meson with the mass $M_{a_1} = 1260~{\rm MeV}$,
the $f_1$ meson, with the mass $M_{f_1}=1285~{\rm MeV}$
and the $f_1^*$ meson with the mass $M_{f_1^*} = 1420~{\rm MeV}$.
The contribution of these mesons to $a_\mu^{\rm lbl}$ is 
cut off  at the scales defined by their masses. 
This suggests that the pole-like singularities in Eq.(\ref{amplpole})
are shifted from zero to the masses of  the corresponding pseudovector and
vector mesons. We also remind the reader that
$\phi^{(a)}_{T}(q_{1}^{2}, q_{2}^{2})$   
describes the form factor for the 
$\gamma^* \gamma^* \to a_1(f_1)$ transition.
Shifting all the poles by the same amount, i.e., 
neglecting mass differences, we get
the simplest possible model 
consistent with perturbative QCD constraints Eq.(\ref{Tcon}),
\ba
&&\phi_{T}^{(a)}(q_1^{2},q_2^{2}) = -\frac{4}{(q_1^2+M^2)(q_2^2+M^2)}\,,
\nonumber\\[1mm]
&& w_T^{(a)}(q) = \frac{1}{q^2+M^2}.
\label{simple}
\ea
This implies, in particular, that we do not distinguish between different isospin 
channels. 

Although this model is not very realistic, we can use it to derive 
a simple analytic result which will help us to exhibit the  dependence 
on the mass scale $M$. Assuming that $M \gg m_\mu$, we 
compute the contribution 
of the pseudovector meson to $a_\mu$ and obtain:
\ba
a_\mu^{\rm PV} &\!\!\! =&\! \!\left ( \frac{\alpha}{\pi} \right)^3 
\!\frac{m_{\mu}^2}{M^2} 
N_c {\rm Tr}\, [{\hat Q}^{4}] \left [   
\frac{71}{192}\!+\frac{81}{16}S_2-\!\frac{7 \pi^2}{144}
\right ] 
\nonumber \\[1mm]
&\! \approx&\! 1010~\frac{m_\mu^2}{M^2} \times 10^{-11}\,,
\label{eqmodel}
\ea
where $S_2 = 0.26043$.
Using $M = 1300~{\rm GeV}$  as an example, 
we obtain  $a_\mu^{\rm PV} = 7 \times 10^{-11}$.

There are two comments we would like to make about this result.
First, we compare it to the existing  estimates 
of the pseudovector meson contribution \cite{bijnes,klbl}. 
In those references, the results 
$2.5 \times 10^{-11}$ and $1.7 \times 10^{-11}$ have been 
obtained. We have checked that the difference between our result 
Eq.(\ref{eqmodel}) and the results of \cite{bijnes,klbl} 
can be 
explained by the absence of the form factor 
for the $\gamma^* \gamma h$ interaction vertex in our model; when 
such a form factor is introduced, 
our result decreases to 
$2.6 \times 10^{-11}$, in good agreement with the 
estimates in \cite{bijnes,klbl}.  

Also, we note that the result Eq.(\ref{eqmodel}) exhibits   
strong sensitivity 
to the mass of the pseudovector meson and the mass parameter 
in the form factor. If we associate the mass scale $M$ in 
Eq.(\ref{eqmodel}) with the mass of the $\rho$-meson, the 
result increases roughly by a factor $4$ and becomes $a_\mu^{\rm pv}
\sim 28 \times 10^{-2}$. Because of the strong sensitivity to 
the mass parameter, we have 
to  introduce a more sophisticated model accounting for the 
mass differences in different isospin channels.

Let us start with the isovector function $w_{T}^{(3)}$. 
This function describes the triangle amplitude that involves
the isovector axial current, the virtual photon and the  
soft photon.
We expect therefore 
that $w_T^{(3)}(q_3^{2})$ should contain two poles 
with respect to $q_3^2$: the first 
one, that corresponds to the $a_{1}(1260)$ pseudovector meson 
and the second one, that corresponds to the vector mesons $\rho, \omega$,
thereby reflecting the properties of the virtual photon.  
Such a model was constructed in Ref.\cite{CMV} where it was required
that, for large values of $q^2$,
 the equality $w_L(q^2) = 2w_T(q^2)$,
  remains valid through  ${\cal O}(q^{-4})$ terms. Such 
a requirement leads to
 \be
 w_T^{((3))}(q_3^{2})\!=\! \frac{1}{m_{a_1}^2 - m_\rho^2}
\left [\frac{m_{a_1}^2 - m_\pi^2}{q^2 + m_\rho^2} 
- \frac{m_\rho^2 - m_\pi^2}{q^2 + m_{a_1}^2} 
 \right],
 \ee
where we do not distinguish between the masses of $\rho$ and $\omega$ mesons.
Correspondingly, the form factor $ \phi_T^{((3))}(q_1^{2}, q_{2}^{2})$ becomes
\be
\phi^{(3)}_{T}(q_1^{2},q_2^{2}) 
 = -\frac{4}{(q_1^2+m_\rho^2)(q_2^2 + m_\rho^2)}\,.
 \ee

For the isoscalar pseudovector mesons $f_{1}(1285)$ and $f_{1}(1420)$ we 
assume the ``ideal'' mixing similar to $\omega$ and $\phi$;
 this assumption is consistent with 
experimental data for decays of these resonances. Then, 
instead of the hypercharge 
and the ${\rm SU}(3)$ singlet weights $W^{(8)}$ and $W^{(0)}$, we use 
\be
W^{(u+d)}=\frac{25}{36}\,,\quad  W^{(s)}=\frac{1}{18},
\ee
and the following expressions for the corresponding functions 
$w_{T}$ and $\phi_T\,$:
\begin{widetext}
\ba
&& w_T^{(u+d)}(q^{2}) = \frac{1}{m_{f_1}^2 - m_\omega^2}
\left [\frac{m_{f_1}^2 - (m_\eta^2/5)}{q^2 + m_\omega^2} 
- \frac{m_\omega^2 - (m_\eta^2/5)}{q^2 + m_{f_1}^2} 
 \right ], \qquad \phi_T^{(u+d)}(q_1^{2},q_2^{2}) 
 = -\frac{4}{(q_1^2+m_\omega^2)(q_2^2 + m_\omega^2)}, \nonumber \\
&& w_T^{(s)}(q^{2}) = \frac{1}{m_{f_1^*}^2 - m_\phi^2}
\left [\frac{m_{f_1^*}^2 + m_{\eta}^2}{q^2 + m_\phi^2} 
- \frac{m_\phi^2 + m_{\eta}^2}{q^2 + m_{f_1^*}^2} 
 \right ],\qquad \qquad\phi_T^{(s)}(q_1^{2},q_2^{2}) 
 = -\frac{4}{(q_1^2+m_\phi^2)(q_2^2 + m_\phi^2)}. 
\label{ffactors}
\ea
\end{widetext}
Note, that these refinements of the simple expression for 
the function $w_T$ in Eq.(\ref{simple}) make the 
effective mass of the pseudovector meson lower.
This leads to the increase in $a_\mu^{\rm PV}$ as compared to 
Eq.(\ref{eqmodel}).
We  obtain the following 
estimate:
\be
a_\mu^{\rm PV} =  (5.7 + 15.6 + 0.8) \times 10^{-11} 
=22 \times 10^{-11},
\label{eqpv0}
\ee
where  the three terms displayed separately 
are due to the  isovector, $u+d$  and $s$ exchanges respectively.

To check the stability of the model, we consider an opposite 
case for the mixing, assuming that $f_{1}$ is a pure octet  and
and $f_{1}^{*}$ is  an ${\rm SU}(3)$ singlet meson. 
The estimate for $a_\mu^{\rm PV}$ then becomes
\be
a_\mu^{\rm PV} = \left (5.7 + 1.9 + 9.7 \right ) \times 10^{-11}
 = 17 \times 10^{-11}.
\label{eqpv}
\ee
We see that the ${\rm SU}(3)$-singlet contribution is  significant, 
in spite of the fact that the corresponding masses are the largest. The reason 
for such a behavior is a stronger coupling of the ${\rm SU}(3)$-singlet 
meson to two photons.  We see also that in spite of 
a very strong redistribution between the 
different ${\rm SU}(3)$ channels, the final result for the pseudovector 
contribution is relatively stable against such variations of the 
model.

We use the result for the pseudovector 
contribution in Eq.(\ref{eqpv0}) in our final estimate of 
$a_\mu^{\rm lbl}$ assigning $\pm 5 \times 10^{-11}$ as an error
estimate.

\section{The anatomy of the pion box contribution}

In this Section we make a few comments concerning 
another contribution to $a_\mu^{\rm lbl}$ frequently 
considered in the literature, the so-called pion box contribution.
This contribution is peculiar because, being 
independent of the number of colors  $N_c$, 
it is enhanced by the other potentially large parameter, 
the small value of the pion mass relative to the scale 
of chiral symmetry breaking $\sim 1~{\rm GeV}$.

The results for the pion box contribution to $a_\mu^{\rm lbl}$ were
obtained in \cite{bijnes,klbl}; they are 
$a_\mu^{\rm pion} = -4.5(8.5) \times 10^{-11}$ in \cite{klbl} and 
$a_\mu^{\rm pion} = -19(5) \times 10^{-11}$ in \cite{bijnes}. The difference 
between the two results is attributed to a different treatment of subleading 
terms in the chiral expansion;
while the vector meson dominance (VMD) model is used in \cite{bijnes} to 
couple photons to pions, the so-called hidden local symmetry (HLS)  model 
is used in \cite{klbl}.\footnote{
The claim in \cite{einhorn} and \cite{klbl} that 
the VMD model violates the Ward identities for the $\gamma^*\gamma^*\pi\pi$ 
amplitude is not correct, if the VMD is implemented in the standard 
way, by introducing the factor 
$(M^2g_{\mu\nu}+q_\mu q_\nu)/(M^2+q^2)$ for each photon 
in any interaction vertex. The Ward identities, discussed in 
\cite{klbl}, are then automatically satisfied.}
 Although the smallness of $a_\mu^{\rm pion}$ 
shows that the chiral enhancement is not efficient for $a_\mu^{\rm lbl}$, 
the strong  sensitivity of the final 
result to the {\it particular method} of including  
heavier resonances
suggests that the chiral expansion 
per se may not be a reliable tool for this problem. If this is true, 
the natural question is to what extent the above estimates 
of the pion box  can be trusted.
With this question  
in mind, we investigated an ``anatomy'' of this contribution based 
on the analytic calculation of $a_\mu^{\rm pion}$
in the framework of the HLS model.

The logic which is behind the use of the chiral expansion 
to estimate subleading ${\cal O}(N_c^{0})$ 
contributions to $a_\mu^{\rm lbl}$ is 
as follows. If the pion box contribution  
to $a_\mu$ is determined  by small values of virtual momenta, 
comparable to the masses of  muon and pion, we can
compute  it by using  chiral perturbation 
theory. The leading term in the chiral expansion delivers
a parametrically enhanced   contribution $(\alpha/\pi)^3 (m_\mu/m_\pi)^2$
to $a_\mu^{\rm lbl}$, which can be derived from 
the scalar QED Lagrangian for the pions:
\be
{\cal L} = D_\mu \pi D^{\mu} \pi^{*} - m_\pi^2 |\pi|^2. 
\label{eq0}
\ee
Here $D_\mu = \partial_\mu - ieA_\mu$ is the covariant derivative
and 
$\pi$ is the pion field.
The Lagrangian Eq.(\ref{eq0}) is the leading term in the 
effective chiral Lagrangian and hence 
the terms neglected in Eq.(\ref{eq0}),
are suppressed by the square of the ratio of the pion mass to the scale 
of the chiral symmetry breaking. Numerically, these 
corrections are expected to be small since $m_\pi^2/M_\rho^2 \sim 0.04$ and
$ m_\pi^2/(4\pi f_\pi)^2 \sim 0.025$; therefore, they 
 should not change the 
scalar QED prediction by more than a few per cent.

It is then puzzling that the 
results available in the literature exhibit drastically 
different behavior. Existing calculations show that the 
scalar QED contribution is reduced by a factor from 
three \cite{bijnes} to ten \cite{klbl}, when  subleading 
terms in the chiral expansion are included.
Hence, the results for the pion box contributions 
existing in the literature tell us that the chiral expansion 
for this contribution does not work.  In order to identify the 
reason for that, we  computed several terms of the expansion 
in $m_\pi/M_\rho$ in the framework of the HLS model. 
Comparing the magnitude of the subsequent terms in the expansion, 
we can  determine the rate of convergence of the chiral 
expansion and estimate the typical virtual momentum in the pion 
box diagram.

As we demonstrate below, the  typical virtualities in the pion box diagram 
are approximately $4m_\pi$ which leads to a slow convergence of 
the chiral  expansion and explains, to a certain extent,  
a very strong cancellation between the leading order scalar 
QED result and the first $m_\pi^2/M_\rho^2$ correction. 
The remaining terms in the chiral expansion are smaller (although 
not negligible).  

Large value of typical virtualities brings in another problem 
with the scalar QED model Eq.(\ref{eq0}) and its  modifications 
based on the VMD. Since fairly large virtualities are involved, 
one might wonder about the quality of the model for asymptotically
large values of $q$. To see that the model fails relatively 
early, we can consider the deep inelastic scattering of 
a virtual photon with large value of $q^2$, on a pion. The Lagrangian 
(\ref{eq0}) then implies the  dominance of the longitudinal structure function,
while QCD predicts the opposite.
Modifying the scalar QED Lagrangian Eq.(\ref{eq0}) to accommodate 
the VMD  either directly or through the HLS model, 
does not fix this problem, since only
 an overall factor $(M^2_\rho/(M^2_\rho + q^2))^2$ is introduced
in the imaginary part of the forward scattering amplitude. 
This mismatch implies that the 
models that we use to compute the pion box contribution 
become unreliable {\em very rapidly} once the energy 
scale of the order of the $\rho$-meson mass is passed. Since 
$4 m_\pi$ is only marginally smaller than $M_\rho$, it is hard 
to tell how big a mistake we make by ignoring the fact 
that our hadronic model has incorrect asymptotic behavior.

The  above considerations  suggest that while it is most 
likely that the pion box contribution to $a_\mu$ is relatively small,
as follows from a strong cancellation of the two first terms in the 
chiral expansion, 
the precise value of this contribution 
is impossible to obtain, using  simple VMD and the like models.

We now perform an analytic calculation of the pion box contribution
to $a_\mu$ and demonstrate that the 
typical loop momenta in the pion box 
amplitude are relatively large.
For the analytic calculation, we use
the HLS model  \cite{klbl} to describe
low-energy hadron-photon interactions.
 From a computational point of view, we have to deal with three-loop
diagrams  that involve three distinct scales: the mass of the muon 
$m_\mu$, the mass of the pion $m_\pi$ and the mass of the $\rho$ meson
$M_\rho$. Because the masses of the muon and the pion are close, 
one can treat them as almost equal and construct an  expansion in their 
mass difference; this reduces the problem to two-scale diagrams. 
As the next step, one  constructs the expansion in $m_\mu/M_\rho$, 
using the theory of asymptotic expansions for Feynman diagrams
(see  \cite{Smirnov:pj} for a review).

As it turns out, there are twelve different momenta 
regions to be considered; 
the two limiting cases are a) all the loop momenta are much 
smaller than the mass of the $\rho$-meson
and  b) all the loop momenta are comparable to the mass of the 
$\rho$ meson.
In case a)  one has to compute the three-loop ``on the mass-shell'' diagrams; 
 in case b)  the masses of both muon and pion 
can be neglected and one has to compute the three loop vacuum 
bubble diagrams. Intermediate cases in which some of 
the loop momenta are small and the other are large factorize 
into the product of one- and  two-loop diagrams. The techniques 
needed for such a computation are described in 
Refs.\cite{Melnikov:2000zc,Melnikov:2000qh}.

We now present the result of the  calculation. To do this 
 in a compact form, we introduce the notation 
$\delta = (m_\mu - m_\pi)/m_\pi$ and $L=\ln(M_\rho/m_\pi)$.
We then write:
\be
a_\mu^{\pi} = \left ( \frac{\alpha}{\pi} \right )^3 
\sum \limits_{i=0}^{\infty} f_i(\delta, L) 
\left ( \frac{m_\pi^2}{M_\rho^2 } \right )^i.
\label{eq9}
\ee
The functions $f_{i}(\delta,L)$ for 
$i=0,1,2$ are given in Appendix II.  We have computed 
$f_i(\delta,L)$ for $i$ from $i=0$ to $i=4$ analytically and 
we use those functions below for numerical estimates.
In addition, we use 
$m_\pi=136.98~{\rm MeV}$, $m_\mu=105.66~{\rm MeV}$ and 
$M_\rho=769~{\rm MeV}$. With these input values,  
Eq.(\ref{eq9}) evaluates to:
\ba
&& a_\mu^{\pi}= -0.0058 \left ( \frac{\alpha}{\pi} \right )^3 = 
\left ( -46.37+35.46 + 10.98
\right. \nonumber \\
&& \left. -4.70-0.3+...
\right ) \times 10^{-11} = -4.9(3) \times 10^{-11},
\label{num25}
\ea
where the subsequent terms in Eq.(\ref{num25}) correspond to
the subsequent terms in Eq.(\ref{eq9}).

The feature of the result Eq.(\ref{num25})  which 
has to be  noticed is 
the strong cancellation between the leading and the subleading 
terms in the chiral expansion; the other terms, being non-negligible 
numerically, are certainly smaller.  It is this cancellation 
that ensures the smallness of the final value for the pion pole 
contribution to $a_\mu$. 
We can use Eq.(\ref{num25}) to determine  typical momenta 
virtualities in the pion box contribution.

For simplicity, we  study this question assuming $m_\pi=m_\mu$, 
which implies $\delta=0$ in the formulas for $f_i(\delta,L)$ presented 
in Appendix II. In this limit Eq.(\ref{num25}) becomes:
\be
a_\mu^{\pi}(m_\pi=m_\mu) \approx
\left (-69 + 54 +18 -8 - 1+.. \right ) \times 10^{-11}.
\label{puks}
\ee
We also  assume that the contribution 
to $a_\mu$ can be described by the chiral expansion, with 
the effective scale $\mu$. This scale characterizes 
the typical virtual momentum in the pion box
diagram.  Motivated by  the chiral perturbation theory, 
we make an Anzats:
\be
a_\mu^{\pi}(m_\pi=m_\mu) \approx \left ( \frac{\alpha}{\pi} \right )^3 
\frac{m_\pi^2}{\mu^2}\left (c_1 + c_2 \frac{\mu^2}{M_\rho^2}
+ c_3 \frac{\mu^4}{M_\rho^4}+...\right ). 
\label{puks1}
\ee
We further assume that all the coefficients in the above series 
are numbers of order one.  Setting $c_1=1$ in the above equation, 
we can determine the value of $\mu$ by comparing it with the first 
term in Eq.(\ref{puks}). We obtain $\mu=4.25 m_\pi$. Then, Eq.(\ref{puks1})
becomes:
\be
a_\mu^{\pi}(m_\pi=m_\mu) \approx
\left (
-69+41 c_2  + 24 c_3  + 14c_4+.. \right ) \times 10^{-11},
\ee
which implies that with $c_2 \approx 1.3$, $c_3 \approx 0.75$ and 
$c_4 \approx -0.6$, we can easily fit Eq.(\ref{puks}).  

The above calculation suggests a simple way to understand the magnitude 
of the  the chirally suppressed terms in Eq.(\ref{num25}). 
Since $\mu \approx 4 m_\pi 
\approx 550~{\rm MeV} < M_\rho$, the chiral expansion converges, but 
rather slowly. Therefore, the estimates based on the chiral 
expansion do  make sense {\it in principle}.
A closer look at $f_i(\delta,L)$ 
reveals that these functions contain $\ln(M_\rho/m_\pi)$-enhanced 
 terms.
However, in view of the above argument, the appropriate 
way to write the large logarithms is $\ln(M_\rho/\mu)$; 
doing so, we observe that  ``large'' logarithms become 
rather moderate numerically and every  function $f_i(\delta,L)$ 
is   dominated by constant terms.

We therefore see that the typical virtual momenta in the pion box
contribution are {\it larger} than the mass of the pion by, 
approximately, a factor of $4$. While the chiral expansion 
is still a valid tool for such virtualities, its predictive 
power becomes small. This can be seen from 
Eq.(\ref{num25}), which implies that the final result for 
the pion box contribution to $a_\mu^{\rm lbl}$
is very sensitive to higher order power 
corrections. It is clear that since none of the models, 
be it the HLS model or the VDM model, can claim 
full control over higher order power corrections in the chiral 
expansion, the {\it exact result} for the pion box contribution 
is not very meaningful. However, the fact that the chiral 
expansion is still applicable suggests that 
the strong cancellation between the leading order term and {\it the first}  
subleading  term in the chiral expansion may be  a generic 
feature of QCD.

Therefore, we find it reasonable 
to {\it believe} that the pion box contribution to $a_\mu^{\rm lbl}$
is much smaller than the estimate based on the chirally enhanced 
scalar QED result  for the pions. However, once this point of view
is accepted,  the chiral enhancement looses its power as the theoretical 
parameter and the pion box contribution becomes just one of many 
 ${\cal O}(N_c^{0})$ contributions about which nothing is known 
at present.  Therefore, for the final estimate of $a_\mu^{\rm lbl}$
we  use
\be
a_{\mu}^{\rm lbl,N_c^{0}} = 0(10) \times 10^{-11},
\label{eqNc0}
\ee
where the error estimate is clearly subjective.

\section{Conclusions}
In this paper, 
we  revisited the issue of the hadronic 
light-by-light scattering contribution to the muon anomalous 
magnetic moment,  incorporating constraints 
from perturbative QCD in constructing the low-energy  model for 
the light-by-light scattering. To achieve that, 
we  computed the light-by-light scattering 
amplitude at a relatively large photon virtualities in perturbative 
QCD and required that low-energy models for hadronic light-by-light 
scattering have to  interpolate smoothly  between 
small and large values of $q^2$. The minimal large-$N_c$ model 
with such feature contains the pseudoscalar and the pseudovector 
meson exchanges.

Since, by construction, the  hadronic model we use in this 
paper has correct scaling at large values of 
$q^2$, we achieve reasonable  matching 
between the low-energy and the high-energy degrees of freedom  
that contribute to hadronic light-by-light scattering amplitude.
One of the major findings in this paper is the fact that 
too strong a damping of hadronic amplitudes at large values of 
$q^2$ has been used in previous studies 
\cite{knecht,klbl,bijnes} of hadronic light-by-light 
scattering to $a_\mu$. 

It turns out that imposing   correct matching between 
the low- and the high-energy degrees of freedom 
leads to substantial changes in both the pseudoscalar and the pseudovector
contributions, making both of them larger. In a way,
the impact of large-momentum degrees of freedom on $a_\mu$ was 
underestimated in previous analyses. Our final result for 
hadronic light-by-light scattering contribution to $a_\mu$ is:
\be
a_{\mu}^{\rm lbl} = 136(25) \times 10^{-11}.
\label{eqfin}
\ee
The error estimate includes the 
sum of ${\cal O}(N_c^{0})$ error estimate in 
Eq.(\ref{eqNc0}) as well as $15 \times 10^{-11}$ 
as an error estimate for  the sum of the pseudoscalar and the pseudovector 
exchanges. From Eq.(\ref{eqfin}) it is clear that we do not  
claim  significant reduction in the theoretical 
uncertainty, although we believe that it adequately reflects 
our current knowledge of $a_{\mu}^{\rm lbl}$.
On the contrary, we think that the shift in the central value 
is real because it originates from a better matching of the 
low-energy hadronic models and the short-distance QCD.

The result in Eq.(\ref{eqfin}) is approximately $50$ percent 
larger, than the currently accepted estimate $\sim 80(30) \times 10^{-11}$
\cite{knecht,klbl,bijnes}. Note however, that our result 
is closer to another 
recent evaluation of hadronic light-by-light scattering contribution 
to $a_\mu$ \cite{kuraev} where the central value 
$a_\mu^{\rm lbl} =108 \times 10^{-11}$ is  quoted.

Another possible consistency check is to estimate the light-by-light 
scattering contribution as a sum of two terms -- the pion-pole 
contribution to account for low-momentum region and 
the massive quark box contribution to account for large-momentum 
regime. If the quark masses are chosen to be $m_q=300~{\rm MeV}$, 
the result for the quark box contribution is $60 \times 10^{-11}$.
Combining this with the pion pole contribution, 
we get the estimate
$a_\mu^{\rm lbl} \approx 120 \times 10^{-11}$.
Of course, the above consideration is not the proof; 
yet it clearly indicates the tendency of the result for $a_\mu^{\rm lbl}$
to increase once the contribution 
of the large-momentum region is
accounted for in the correct way.

Finally, we note that the new value for hadronic light-by-light 
scattering contribution, Eq.(\ref{eqfin}), brings
the estimate of the muon magnetic moment 
anomaly in the Standard Model and the current experimental value
\cite{g-2recent}
somewhat closer. Using  recent results 
for hadronic vacuum polarization \cite{eidelman}, 
we arrive at:
\be
 a_{\mu}^{{\rm exp}} - a_{\mu}^{{\rm th}} = 
\left \{ 
\begin{array}{cc}
(171 \pm 110) \times 10^{-11} & (1.5~\sigma),\\
(24 \pm 110) \times 10^{-11} & (0.2~\sigma).
\end{array} \right.
\ee
The first result uses $e^+e^-$-data only, while the second 
one uses the $\tau$ data at low energies; the errors in each 
equation are combined in quadratures for compactness.

\vspace*{0.5cm}
{\bf Acknowledgments:} We are grateful 
to M.~Voloshin for helpful discussions.
This research is supported by the DOE
under contracts DE-FG03-94ER-40833 and DE-FG02-94ER408 and 
the Outstanding Junior Investigator Award DE-FG03-94ER40833.

\section*{Appendix I}

In this Appendix we give explicit expressions for the 
light-by-light scattering amplitude in perturbative QCD 
in the kinematics when three photons have non-zero virtualities 
and one of the photons is soft.

The Euclidean amplitude 
\ba
{\cal A}=\epsilon_{1}^{\mu_{1}}\epsilon_{2}^{\mu_{2}}\epsilon_{3}^{\mu_{3}}
\epsilon_{4}^{\mu_4}
{\cal A}_{\mu_{1}\mu_{2}\mu_{3} \mu_4},
\ea
defined in Eq.(\ref{eqcala})
can be expressed through  nineteen gauge-invariant structures:
\begin{widetext}
\ba
{\cal A}&=&G_{1}^{(1,2,3)}
\left\{f f_{1}\right\}\left\{f_{2} f_{3}\right\}
+G_{1}^{(2,3,1)}\left\{f f_{2}\right\}\left\{f_{3} f_{1}\right\}
+G_{1}^{(3,1,2)}\left\{f f_{3}\right\}\left\{f_{1} f_{2}\right\}
\nonumber\\[2mm]
&+&G_{2}^{(1,2,3)}\left\{f \tilde f_{1}\right\}
\left\{f_{2}  \tilde f_{3}\right\}
+G_{2}^{(2,3,1)}
\left\{f \tilde f_{2}\right\}\left\{f_{3} \tilde f_{1}\right\}
+G_{2}^{(3,1,2)}
\left\{f \tilde f_{3}\right\}\left\{f_{1} \tilde f_{2}\right\}
\nonumber\\[2mm]
&+&G_{3}^{(1,2,3)}
\left\{
\eta_{23}f\,f_{1}\,\eta_{23}
\right\}\left\{f_{2}f _{3}\right\}
+G_{3}^{(2,3,1)}
\left\{
\eta_{31}f\,f_{2}\,\eta_{31}
\right\}\left\{f _{3}f _{1}\right\}
+
G_{3}^{(3,1,2)}
\left\{
\eta_{12}f\,f_{3}\,\eta_{12}
\right\}\left\{f_{1}f _{2}\right\}
\nonumber\\[2mm]
&+&\tilde G_{3}^{(1,2,3)}
\left\{\eta_{23}f\,f_{1}\,q_{1}
\right\}\left\{f_{2}\,f_{3}\right\}
+\tilde G_{3}^{(2,3,1)}
\left\{\eta_{31}f\,f_{2}\,q_{2}
\right\} \left\{f _{3}\,f _{1}\right\}
+\tilde G_{3}^{(3,1,2)}
\left\{\eta_{12}f\,f_{3}\,q_{3}
\right\}\left\{f _{1}\,f _{2}\right\}
\nonumber\\[2mm]
&+&G_{4}^{(1,2,3)}
\left\{\eta_{23}f\,f_{1}\,\eta_{23}
\right\}\left\{q_{2}f _{2}\,\eta_{31}\right\}
\left\{q_{3}f _{3}\,\eta_{12}\right\}
+G_{4}^{(2,3,1)}
\left\{\eta_{31}f\,f_{2}\,\eta_{31} \right\}
\left\{q_{3}f _{3}\,\eta_{12}\right\}
\left\{q_{1}f _{1}\,\eta_{23}\right\}
\nonumber\\[2mm]
&+&G_{4}^{(3,1,2)}
\left\{\eta_{12} f\,f_{3}\,\eta_{12}
\right\}
\left\{q_{1}f _{1}\,\eta_{23}\right\}
\left\{q_{2}f _{2}\,\eta_{31}\right\}
\nonumber\\[2mm]
&+&\tilde G_{4}^{(1,2,3)}
\left\{q_1f\,f_{1}\,q_1
\right\}\left\{q_{2}f _{2}\,\eta_{31}\right\}
\left\{q_{3}f _{3}\,\eta_{12}\right\}
+\tilde G_{4}^{(2,3,1)}
\left\{q_2 f\,f_{2}\,q_2
\right\}
\left\{q_{3}f _{3}\,\eta_{12}\right\}
\left\{q_{1}f _{1}\,\eta_{23}\right\}
\nonumber\\[2mm]
&+&\tilde G_{4}^{(3,1,2)}
\left\{q_3 f\,f_{3}\,q_3
\right\}
\left\{q_{1}f _{1}\,\eta_{23}\right\}
\left\{q_{2}f _{2}\,\eta_{31}\right\}
+G_{5}^{(1,2,3)}
\left\{q_{1}f \,q_{3}\right\}
\left\{q_{1}f _{1}\,\eta_{23}\right\}
\left\{q_{3}f _{3}\,\eta_{12}\right\}
\left\{q_{2}f _{2}\,\eta_{31}\right\}.
\label{ampl}
\ea
\end{widetext}
We have introduced the field strength tensor for all of the 
four photons 
$f_i^{\mu\nu}=q_i^{\mu}\epsilon_i^{\nu} - q_i^{\nu}\epsilon_i^{\mu}$ with 
$f_4^{\mu\nu}= f^{\mu\nu}$
and, also, the 
four-vectors $\eta_{ij}=q_i-q_j$.
In  Eq.(\ref{ampl}), we  
view $f_{i}^{\mu\nu}$ as matrices; the curly brackets imply
either traces  of matrix products or convolutions with 
vectors $q_{i},\eta_{ij}$.
For example, $\{q_1 f q_2 \} = q_{1,\mu} f^{\mu \nu} q_{2,{\nu}}$. 
The notations for invariant functions $G_{1-5}$ are introduced for 
compactness. The superscripts denote the arguments of these 
functions, e.g. $G_5^{(1,2,3)} = G_5(q_1^2,q_2^2,q_3^2)$.
The invariant function $G_{5}(q_{1}^{2},q_{2}^{2},q_{3}^{2})$ is
 totally symmetric with respect to the permutations of its arguments; 
the  functions $G_{1,2,3,4}(q_{1}^{2},q_{2}^{2},q_{3}^{2})$ are 
symmetric under the permutation of the last two arguments; the functions
 $\tilde G_{3,4}(q_{1}^{2},q_{2}^{2},q_{3}^{2})$ are 
antisymmetric under the permutation of the last two arguments.

We have computed the above form factors in perturbative QCD where 
the photon-photon interaction is mediated by the loops of massless 
quarks. Our results are:
\begin{widetext}
\ba
&& G_1(s_1,s_2,s_3)=-\left (4s_3s_2^4
-4s_3^3s_2^2+6s_3^2s_1^3-4s_3s_1^4-22s_1^3s_2^2
+28s_3s_2^2s_1^2+26s_3^2s_2^2s_1-11s_2^4s_1+32s_1^2s_2^3
\right. \nonumber \\
&& \left.
-48s_3s_2^3s_1
-2s_2^5+s_3^4s_1+2s_1^4s_2-4s_3s_2s_1^3-32s_3^2s_2s_1^2+32s_3^3s_2s_1
+2s_3^4s_2+s_1^5-4s_3^3s_1^2
\right )\frac{\ln(s_3)}{D^2 s_1 (s_1-s_3-s_2) s_2}
\nonumber \\
&& +\left (s_2^5-4s_2^4s_1+6s_1^2s_2^3+6s_3^3s_1^2+21s_3s_2^4-22s_3^2s_2^3
-22s_3^3s_2^2-26s_3s_2^2s_1^2+56s_3^2s_2^2s_1-4s_1^3s_2^2+21s_3^4s_2
\right. \nonumber \\
&& \left.
+4s_3s_2s_1^3
-26s_3^2s_2s_1^2+s_3^5+s_1^4s_2-4s_3^2s_1^3-4s_3^4s_1
+s_3s_1^4 \right. ) \frac{\ln(s_1)}{D^2 s_3 (s_1-s_3-s_2) s_2}
\nonumber \\
&& -\left (-4s_1^4s_2-32s_3s_2^2s_1^2
+28s_3^2s_2s_1^2+32s_3s_2^3s_1+26s_3^2s_2^2s_1-4s_1^2s_2^3-2s_3^5
+6s_1^3s_2^2-4s_3^2s_2^3+s_2^4s_1+2s_3s_2^4
\right. \nonumber \\
&& \left.
+4s_3^4s_2-11s_3^4s_1
+s_1^5-4s_3s_2s_1^3-48s_3^3s_2s_1-22s_3^2s_1^3+32s_3^3s_1^2
+2s_3s_1^4\right )\frac{\ln(s_2)}{D^2 s_1 s_3 (s_1-s_3-s_2)}
\nonumber \\
&& -2 \left ( 18s_3^2s_2^2+7s_2^3s_1
-7s_1^3s_2+18s_3s_2s_1^2-19s_3s_2^2s_1-4s_3^3s_2+3s_1^2s_3^2
-4s_3s_2^3-7s_1^3s_3+2s_1^4+3s_1^2s_2^2
\right. \nonumber \\
&& \left.
+7s_1s_3^3
-19s_3^2s_2s_1
-5s_3^4-5s_2^4 \right )\frac{J(s_1,s_2,s_3)}{D^2(s_1-s_3-s_2)} 
-4\frac{(-s_2^2-3s_1s_3+s_1^2
-s_3^2-3s_2s_1+2s_3s_2)}{(s_1-s_3-s_2) s_1 D},
\nonumber \\
\nonumber \\
\nonumber \\
&& G_2(s_1,s_2,s_3) = 
-8 \left( 2s_2^2-4s_3s_2+2s_3^2-s_2s_1-s_1s_3-s_1^2 \right 
)\frac{\ln(s_1)}{D^2}
+4 (s_2-s_1-s_3)\left( s_3^2-2s_3s_2
\right. \nonumber \\
&& \left. -2s_1s_3+s_1^2+4s_2s_1
+s_2^2 \right )\frac{\ln(s_2)}{D^2 s_1}
-4(-s_3+s_2+s_1)\left(s_3^2-2s_3s_2+4s_1s_3
+s_2^2+s_1^2
-2s_2s_1 \right )\frac{\ln(s_3)}{D^2s_1}
\nonumber \\
&& -8(s_2^3-s_2^2s_3-s_3^2s_2+s_3^3-2s_1s_2^2+2s_1s_3s_2-2s_1s_3^2
+s_1^2s_2+s_1^2s_3)\frac{J(s_1,s_2,s_3)}{D^2}
-\frac{8}{D},
\nonumber \\
\nonumber \\
\nonumber \\
&& G_3(s_1,s_2,s_3)=
2\left (5s_3^4s_1-10s_3^3s_1^2-26s_3^2s_2^3+22s_3^2s_2s_1^2-20s_3s_2s_1^3
-s_1^4s_2+54s_3^3s_2^2
\right. \nonumber \\
&& \left.
+10s_3^2s_1^3+15s_2^5-47s_2^4s_1
+50s_1^2s_2^3
-38s_3s_2^2s_1^2-18s_1^3s_2^2+20s_3^3s_2s_1-21s_3s_2^4+84s_3s_2^3s_1
\right. \nonumber \\
&& \left.
-62s_3^2s_2^2s_1
-5s_3s_1^4-21s_3^4s_2-s_3^5+s_1^5 \right )
\frac{\ln(s_3)}{s_2D^3}
\nonumber \\
&& -2\left(
5s_3^5s_1+s_3s_1^5-6s_3^5s_2-10s_3^4s_1^2-s_2^6-s_3^6-38s_3^2s_2s_1^3
-38s_3s_2^2s_1^3+22s_3^2s_2^3s_1+22s_3^3s_2^2s_1
\right. \nonumber \\
&& \left.
-27s_3s_2^4s_1
-2s_3s_2s_1^4
+72s_3^3s_2s_1^2-76s_3^2s_2^2s_1^2+72s_3s_2^3s_1^2-27s_3^4s_2s_1+s_1^5s_2
+33s_3^2s_2^4
\right. \nonumber \\
&& \left.
-6s_3s_2^5-52s_3^3s_2^3
+33s_3^4s_2^2+5s_2^5s_1
-10s_1^2s_2^4
-5s_1^4s_2^2-5s_3^2s_1^4+10s_3^3s_1^3+10s_1^3s_2^3 \right )
\frac{\ln(s_1)}{s_2 s_3 D^3}
\nonumber \\
&& +2 \left (
5s_2^4s_1+22s_3s_2^2s_1^2-s_2^5-62s_3^2s_2^2s_1+20s_3s_2^3s_1+s_1^5
-21s_3^4s_2-26s_3^3s_2^2+15s_3^5+10s_1^3s_2^2+84s_3^3s_2s_1
\right. \nonumber \\
&& \left.
-38s_3^2s_2s_1^2
-10s_1^2s_2^3+54s_3^2s_2^3-18s_3^2s_1^3+50s_3^3s_1^2-20s_3s_2s_1^3-s_3s_1^4
-5s_1^4s_2-47s_3^4s_1-21s_3s_2^4 \right ) \frac{\ln(s_2)}{s_3D^3}
\nonumber \\
&& +4 \left (-8s_1^2s_2^3
+18s_1^3s_2^2-48s_3^2s_2^2s_1-4s_3s_2^2s_1^2-9s_3s_2^4-9s_3^4s_2
+3s_2^5+3s_3^5-11s_3s_1^4-4s_3^4s_1+2s_1^5
\right. \nonumber \\
&& \left.
+28s_3s_2^3s_1+18s_3^2s_1^3
-8s_3^3s_1^2+28s_3^3s_2s_1-4s_2^4s_1-11s_1^4s_2+6s_3^3s_2^2-4s_3s_2s_1^3
-4s_3^2s_2s_1^2+6s_3^2s_2^3 \right )\frac{J(s_1,s_2,s_3)}{D^3}
\nonumber \\
&& 
-\frac{8(s_1^2-4s_3^2+8s_3s_2-4s_2^2
+3s_1s_3+3s_2s_1)}{D^2},
\nonumber \\
\nonumber \\
\nonumber \\
&& \tilde G_3(s_1,s_2,s_3)= -2
\left ( -s_3^5s_1-5s_3s_1^5-2s_3^5s_2+5s_3^4s_1^2
-2s_2^6+s_1^6-30s_3^2s_2s_1^3+26s_3s_2^2s_1^3+98s_3^2s_2^3s_1-10s_3^3s_2^2s_1
\right. \nonumber \\
&& \left. -53s_3s_2^4s_1+10s_3s_2s_1^4+56s_3^3s_2s_1^2-114s_3^2s_2^2s_1^2
+16s_3s_2^3s_1^2-27s_3^4s_2s_1-7s_1^5s_2-4s_3^2s_2^4+6s_3s_2^5-4s_3^3s_2^3
+6s_3^4s_2^2
\right. \nonumber \\
&& \left. 
-7s_2^5s_1
+37s_1^2s_2^4+28s_1^4s_2^2+10s_3^2s_1^4-10s_3^3s_1^3
-50s_1^3s_2^3 \right )\frac{\ln(s_3)}{s_1 s_2 D^3 }
-2(s_2-s_3)\left (
-s_2^5+5s_2^4s_1-21s_3s_2^4+24s_3s_2^3s_1
\right. \nonumber \\
&& \left. 
-10s_1^2s_2^3
+22s_3^2s_2^3-106s_3^2s_2^2s_1+10s_1^3s_2^2+10s_3s_2^2s_1^2
+22s_3^3s_2^2-21s_3^4s_2-5s_1^4s_2-8s_3s_2s_1^3+24s_3^3s_2s_1
\right. \nonumber \\
&& \left. 
+10s_3^2s_2s_1^2+5s_3^4s_1-s_3^5-10s_3^3s_1^2
+s_1^5+10s_3^2s_1^3-5s_3s_1^4
\right. ) \frac{\ln(s_1)}{s_2 s_3D^3 }
+2\left (-7s_3^5s_1-7s_3s_1^5+6s_3^5s_2+37s_3^4s_1^2
\right. \nonumber \\
&& \left. 
+s_1^6-2s_3^6+26s_3^2s_2s_1^3-30s_3s_2^2s_1^3-10s_3^2s_2^3s_1
+98s_3^3s_2^2s_1-27s_3s_2^4s_1+10s_3s_2s_1^4+16s_3^3s_2s_1^2
-114s_3^2s_2^2s_1^2
\right. \nonumber \\
&& \left. 
+56s_3s_2^3s_1^2-53s_3^4s_2s_1-5s_1^5s_2
+6s_3^2s_2^4-2s_3s_2^5-4s_3^3s_2^3-4s_3^4s_2^2-s_2^5s_1
\right. \nonumber \\
&& \left. 
+5s_1^2s_2^4+10s_1^4s_2^2+28s_3^2s_1^4-50s_3^3s_1^3
-10s_1^3s_2^3 \right )\frac{\ln(s_2)}{s_1 s_3 D^3}
-4(s_2-s_3) \left(-5s_2^4+14s_2^3s_1-4s_3s_2^3-12s_1^2s_2^2
\right. \nonumber \\
&& \left. 
-26s_3s_2^2s_1
+18s_3^2s_2^2+28s_3s_2s_1^2+2s_1^3s_2-4s_3^3s_2-26s_3^2s_2s_1
+s_1^4-12s_1^2s_3^2+14s_1s_3^3-5s_3^4+2s_1^3s_3
\right )\frac{J(s_1,s_2,s_3)}{D^3}
\nonumber \\
&& -\frac{8(s_2-s_3)}{
(-s_2^2+2s_3s_2-3s_2s_1-s_3^2+4s_1^2-3s_1s_3) s_1 D^2},
\nonumber \\
\nonumber \\
\nonumber \\
&& G_4(s_1,s_2,s_3)= 
-4\left (-s_3^4-10s_1^3s_2+8s_3s_2^3-2s_1^3s_3+2s_1s_3^3+s_1^4-17s_2^4
+48s_3s_2s_1^2-78s_3s_2^2s_1
\right. \nonumber \\
&& \left.
-14s_3^2s_2s_1-24s_3^3s_2+26s_2^3s_1
+34s_3^2s_2^2 \right )\frac{\ln(s_3)}{s_2 (s_1-s_3-s_2) (s_2-s_3) D^3}
-4 \left (
-s_2^4-8s_3s_2^3+2s_2^3s_1+18s_3^2s_2^2
\right. \nonumber \\
&& \left. 
-38s_3s_2^2s_1-8s_3^3s_2
+48s_3s_2s_1^2-2s_1^3s_2-38s_3^2s_2s_1-s_3^4+s_1^4-2s_1^3s_3
+2s_1s_3^3 \right )\frac{\ln(s_1)}{D^3 s_3 (s_1-s_3-s_2) s_2}
\nonumber \\
&& +4 \left (
s_1^4-78s_3^2s_2s_1-s_2^4-2s_1^3s_2+2s_2^3s_1-10s_1^3s_3+26s_1s_3^3
-17s_3^4+34s_3^2s_2^2-24s_3s_2^3+8s_3^3s_2
\right. \nonumber \\
&& \left. +48s_3s_2s_1^2
-14s_3s_2^2s_1 \right )\frac{\ln(s_2)}{(s_1-s_3-s_2)(s_2-s_3)s_3 D^3}
+24 \left (s_2^3-s_3s_2^2+s_2^2s_1+10s_3s_2s_1-5s_1^2s_2-s_3^2s_2
\right. \nonumber \\
&& \left.
+s_3^2s_1
 +s_3^3+3s_1^3-5s_3s_1^2 \right ) 
\frac{J(s_1,s_2,s_3)}{D^3 (s_1-s_3-s_2)}
+\frac{8(s_2^2-2s_2s_1+8s_3s_2+s_1^2+s_3^2-2s_1s_3)}{D^2s_3(s_1-s_3-s_2)s_2},
\nonumber \\
\nonumber \\
\nonumber \\
&& \tilde G_4(s_1,s_2,s_3)= 
4 \left (-2s_3^4s_1-22s_1^3s_2^2+4s_1^2s_2^3
+4s_2^4s_1-4s_3s_1^4+22s_3^2s_2^2s_1
-26s_3^2s_2s_1^2+48s_3s_2s_1^3-2s_3s_2^4
\right. \nonumber \\
&& \left.
-26s_3s_2^2s_1^2+11s_1^4s_2
+2s_3^3s_2^2+4s_3^3s_1^2+2s_1^5-32s_3^3s_2s_1-s_3^4s_2
+s_2^5+8s_3s_2^3s_1 \right )\frac{\ln(s_3)}{s_1 (s_1-s_2+s_3) D^3 s_2^2}
\nonumber \\
&& -4 \left (
2s_3s_1^4+4s_3^4s_1+26s_3^2s_2^2s_1+2s_2^4s_1-s_2^5-4s_3^2s_1^3
-2s_3^5-48s_3^3s_2s_1+26s_3^2s_2s_1^2
-22s_3s_2^2s_1^2-8s_3s_2^3s_1
\right. \nonumber \\
&& \left.
+22s_3^3s_2^2-11s_3^4s_2-2s_1^3s_2^2+32s_3s_2s_1^3-4s_3^2s_2^3
-4s_3s_2^4+s_1^4s_2
\right ) \frac{\ln(s_1)}{(s_1-s_2+s_3)D^3 s_3 s_2^2}
\nonumber \\
&& 
+4 \left (-2s_1^4s_2-2s_3^4s_2+52s_3^2s_2s_1^2+s_1^5+21s_3s_1^4
+21s_3^4s_1+2s_3^2s_2^3-s_3s_2^4-s_2^4s_1-8s_3s_2^3s_1-22s_3^3s_1^2
\right. \nonumber \\
&& \left.
-22s_3^2s_1^3
+2s_1^2s_2^3-12s_3s_2^2s_1^2-12s_3^2s_2^2s_1
+s_3^5 \right )\frac{\ln(s_2)}{s_2s_1(s_1-s_2+s_3)D^3s_3}
\nonumber \\
&& +8 \left (17s_3^2s_2s_1+17s_3s_2s_1^2-5s_3^3s_2-5s_1^3s_2+2s_2^4
-18s_1^2s_3^2+s_3s_2^3+5s_3^4+4s_1^3s_3+4s_1s_3^3+5s_1^4
\right. \nonumber \\
&& \left.
-3s_3^2s_2^2-3s_1^2s_2^2+s_2^3s_1-10s_3s_2^2s_1 \right )
\frac{J(s_1,s_2,s_3)}{s_2 (s_1-s_2+s_3) D^3 }
\nonumber \\
&& + \frac{8(s_2^3s_1-2s_1^2s_2^2+s_1^3s_2+s_3s_2^3-2s_3^2s_2^2+s_3^3s_2
+2s_1s_3^3-4s_1^2s_3^2+2s_1^3s_3+5s_3^2s_2s_1+5s_3s_2s_1^2
-2s_3s_2^2s_1)}{s_2^2 s_3 D^2 (s_1-s_2+s_3)s_1}:
\nonumber \\
\nonumber \\
\nonumber \\
&&G_5(s_1,s_2,s_3)= {\cal R}_1(s_1,s_2,s_3) 
+ {\cal R}_1(s_3,s_2,s_1) 
 +{\cal R}_1(s_1,s_3,s_2) +{\cal R}_2(s_1,s_2,s_3);
\nonumber \\
\nonumber \\
&& {\cal R}_1(s_1,s_2,s_3) = 8 \left(s_1^9+s_3^9+s_2^9-16s_1^5s_3^4
-513s_1^5s_3s_2^3-181s_2^4s_1^5+1737s_1^5s_3^2s_2^2
\right. \nonumber \\ [1.5mm]
&& \left.
-515s_1^5s_3^3s_2
-16s_3^5s_2^4-10s_3^7s_2^2-s_3s_2^8+26s_3^3s_2^6+26s_3^6s_2^3
-16s_3^4s_2^5-s_3^8s_2-s_3s_1^8+26s_3^6s_1^3-10s_3^2s_1^7
\right. \nonumber \\ [1.5mm]
&& \left.
-10s_3^7s_1^2+26s_3^3s_1^6-16s_3^5s_1^4-10s_3^2s_2^7-s_1s_3^8
-125s_1^3s_3^5s_2
-1544s_1^3s_3^2s_2^4-453s_1^2s_3^6s_2-446s_1^2s_3s_2^6
\right. \nonumber \\ [1.5mm]
&& \left.
+1862s_1^2s_3^5s_2^2+1737s_1^2s_3^2s_2^5-990s_1^2s_3^3s_2^4
-1645s_1^2s_3^4s_2^3+177s_1s_2^7s_3-183s_1s_2^6s_3^2+207s_1s_2s_3^7
\right. \nonumber \\ [1.5mm]
&& \left.
+909s_1s_2^4s_3^4-125s_1s_2^3s_3^5-515s_1s_2^5s_3^3-453s_1s_2^2s_3^6
+251s_1^3s_2^6-55s_1^2s_2^7-16s_1s_2^8-181s_1^4s_2^5-55s_1^7s_2^2
\right. \nonumber \\ [1.5mm]
&& \left.
+251s_1^6s_2^3+1566s_1^4s_3s_2^4
-990s_1^4s_3^3s_2^2+909s_1^4s_3^4s_2-1544s_1^4s_3^2s_2^3
-1645s_1^3s_3^4s_2^2-513s_1^3s_3s_2^5+3550s_1^3s_3^3s_2^3
\right. \nonumber \\ [1.5mm]
&& \left.
-16s_1^8s_2+177s_1^7s_3s_2-446s_1^6s_3s_2^2
-183s_1^6s_3^2s_2 \right )
\frac{\ln(s_3)}{
D^4D_1 s_1 s_2 (s_1-s_3)(s_2-s_3)}
\nonumber \\ [1.5mm]
&&{\cal R}_2(s_1,s_2,s_3) = 
-16 \left (-54s_3^5s_1-54s_3s_1^5-54s_3^5s_2-45s_3^4s_1^2+29s_2^6
+29s_1^6+29s_3^6-288s_3^2s_2s_1^3
\right. \nonumber \\[1.5mm]
&& \left. 
-288s_3s_2^2s_1^3-288s_3^2s_2^3s_1
-288s_3^2s_2^3s_1
 -288s_3^3s_2^2s_1+342s_3s_2^4s_1+342s_3s_2s_1^4-288s_3^3s_2s_1^2
\right. \nonumber \\[1.5mm]
&& \left.
+666s_3^2s_2^2s_1^2-288s_3s_2^3s_1^2
+342s_3^4s_2s_1-54s_1^5s_2
-45s_3^2s_2^4-54s_3s_2^5+140s_3^3s_2^3-45s_3^4s_2^2-54s_2^5s_1
-45s_1^2s_2^4
\right. \nonumber \\[1.5mm]
&& \left.
-45s_1^4s_2^2-45s_3^2s_1^4+140s_3^3s_1^3
+140s_1^3s_2^3 \right )
\frac{J(s_1,s_2,s_3)}{D^4D_1}
\nonumber \\ [1.5mm]
&& -16
\left (s_1^6+58s_3s_2s_1^4-9s_1^4s_2^2-9s_3^2s_1^4+16s_3^3s_1^3
-58s_3s_2^2s_1^3+16s_1^3s_2^3-58s_3^2s_2s_1^3-9s_3^4s_1^2
\right. \nonumber \\ [1.5mm]
&& \left.
+134s_3^2s_2^2s_1^2-58s_3^3s_2s_1^2-9s_1^2s_2^4-58s_3s_2^3s_1^2
+58s_3s_2^4s_1+58s_3^4s_2s_1-58s_3^2s_2^3s_1-58s_3^3s_2^2s_1+s_3^6
\right. \nonumber \\ [1.5mm]
&& \left.
-9s_3^2s_2^4-9s_3^4s_2^2+s_2^6+16s_3^3s_2^3
\right )\frac{1}
{D^3D_1s_1 s_2 s_3}.
\ea
In the formulas above 
\ba
&& D=s_3^2+s_1^2+s_2^2-2s_1s_2-2s_1s_3-2s_2s_3\,,
\\[1mm]
&& D_1=(s_2+s_1-s_3)(s_1-s_3-s_2)(s_1+s_3-s_2),
\nonumber 
\ea
 and the function $J(s_1,s_2,s_3)$ is 
defined through
\be
J(s_1,s_2,s_3) =  \int \frac{{\rm d}^4 l}{4\pi^2}
\frac{1}{l^2(l+q_1)^2(l-q_3)^2},
\ee
where $q_i^2=s_i$.
The  function $J(s_1,s_2,s_3)$ is symmetric w.r.t. to all its arguments. 
The explicit expression for this function in terms of the polylogarithms
of rank two can be found in \cite{magic}.

\section*{Appendix II}

Below we give the results for the functions $f_{i}(\delta,L)$ for 
$i=0,1,2$ introduced in Eq.(\ref{eq9}):
\ba
&& f_0(\delta,L) = 
  - \frac{11}{72} - \frac{16}{3}a_4 + \frac{11}{36}\zeta_3 \pi^2 
- \frac{1}{6}\zeta_3 - \frac{5}{4}\zeta_5 
+ 12\pi^2\ln 2
+ \frac{2}{9}\pi^2 \ln^2 2 - \frac{1925}{216}\pi^2 + \frac{31}{540}\pi^4 
 - \frac{2}{9}\ln^4 2 
\nonumber \\
&&      + \delta \left ( \frac{1}{36} + 16a_4 - \frac{5}{18}\zeta_3 \pi^2 
 + 6 \zeta_3 + \frac{5}{6}\zeta_5 - 12 \pi^2 \ln 2 
 - \frac{2}{3}\pi^2 \ln^2 2  + \frac{943}{108} \pi^2 - \frac{79}{540}\pi^4 
  + \frac{2}{3}\ln^4 2 \right )
\nonumber \\
&&
       + \delta^2 \left ( \frac{1}{72} - \frac{64}{3}a_4 
 + \frac{5}{12}\zeta_3\pi^2 - \frac{197}{24}\zeta_3 - \frac{5}{4}\zeta_5
          + \frac{47}{4}\pi^2\ln 2 
+ \frac{8}{9}\pi^2\ln^2 2 
 - \frac{479}{54}\pi^2 + \frac{113}{540}\pi^4 
          - \frac{8}{9}\ln^4 2 \right )
\nonumber \\
&&        +\delta^3 \left (
 \frac{7}{54}+\frac{161}{18}\zeta_3-\frac{104}{9}\pi^2\ln 2
 +\frac{5905}{648}\pi^2
        -\frac{55}{216}\pi^4
+24a_4-\pi^2\ln^2 2 +\ln^4 2 
 -\frac{5}{9}\zeta_3\pi^2+\frac{5}{3}\zeta_5 \right ).
\ea\\[-8mm]
\ba
&& f_1(\delta, L) = 
        \frac{3}{2}L^2  + \left (  \frac{13}{4}- \frac{2 \pi^2}{3} \right ) L 
 + \frac{29}{9} + \frac{40}{3}a_4 
- \frac{27}{8}S_2 
+ \frac{4\zeta_3 \pi^2}{3} + \frac{67\zeta_3}{6} 
 - \frac{20 \zeta_5}{3}
\nonumber \\
&&           - \frac{5 \pi^2}{9}\ln^2 2 
- \frac{34\pi^2}{27}  - \frac{31\pi^4}{216} + \frac{5}{9}\ln^4 2 
+ \delta \left (
3L^2  - \left ( \frac{2}{3}\pi^2 + \frac{1}{2} \right )L 
  - \frac{97}{36} - \frac{80}{3}a_4 - \frac{27}{4}S_2 
\right. \nonumber \\
&& \left.
  - \frac{37}{2}\zeta_3 + 
        \frac{10}{9}\pi^2\ln^2 2 + \frac{127}{54}\pi^2 
  + \frac{11}{60}\pi^4 - \frac{10}{9}\ln^4 2 \right  )
 + \delta^2
 \left (   \frac{3}{2}L^2 
  - \left ( \frac{1}{2}\pi^2 - \frac{3}{4} \right )L
+ \frac{115}{24} + \frac{56}{3}a_4 - \frac{27}{8}S_2 
\right. \nonumber \\
&& \left.
+ \frac{263\zeta_3}{24} 
- \pi^2\ln 2 - \frac{7\pi^2}{9}\ln^2 2 
- \frac{53\pi^2}{72} - \frac{13\pi^4}{216} + \frac{7}{9}\ln^4 2 
 \right ).
\ea\\[-8mm]
\ba
&& f_2(\delta,L) = 
          6L^3 
       - \frac{329}{36}L^2 
         + \left (\frac{259\pi^2}{72} -6 \zeta_3 - \frac{14813}{432}
- \frac{27}{8}S_2 \right )L 
  - \frac{40915}{1728} + 16a_4 
        - \frac{45}{8}S_2 - \frac{783}{32}S_2^2 
  - \frac{\pi^3}{36\sqrt{3}} 
\nonumber \\
&&  
+ \frac{1547}{36}\zeta_3
- \frac{2}{3}\pi^2\ln^2 2 
       - \frac{217}{162}\pi^2 
        + \frac{313}{4320}\pi^4 + \frac{2}{3}\ln^4 2 
    + \delta \left (  
         \frac{110}{9}L^3       - \frac{130}{9}L^2
- \left ( 12\zeta_3 - \frac{125\pi^2}{36} + \frac{5239}{108}
\right. \right. \nonumber \\
&& \left. \left. 
- \frac{27}{4}S_2 \right ) L
- \frac{30175}{1296} - 32a_4 
          - \frac{45}{2}S_2 - \frac{459}{8}S_2^2 
      - \frac{\pi^3}{18\sqrt{3}} 
+ \frac{673}{18}\zeta_3 
+ \frac{4}{3}\pi^2\ln^2 2
  + \frac{160}{81}\pi^2 
        + \frac{101}{216}\pi^4 - \frac{4}{3}\ln^4 2 \right ).
\ea
Here, $L=\ln(M_\rho/m_\pi)$, $\delta=(m_\mu-m_\pi)/m_\pi$, 
$\zeta_n$ are the Riemann  zeta functions, $a_4  = {\rm Li}_4(1/2)$ and 
$S_2=0.260434137632161$.
\end{widetext}

\end{document}